\documentclass[11pt]{article}

\newlength{\vshift}
\newlength{\hshift}
\usepackage{amssymb,amsopn}

\renewcommand{\theequation}{\thesection.\arabic{equation}}
\newcommand{\initiate}{\setcounter{equation}{0}}

\def\nn{\nonumber }

\def\be{\beta}
\def\a{\alpha}

\def\g{\gamma}

\def\ds{\stackrel{\star}{,}}

\def\slash{{\rlap /}}

\def\nn{\nonumber}
\def\be{\begin{equation}}             \def\ee{\end{equation}}
\def\ba#1{\begin{array}{#1}}          \def\ea{\end{array}}
\def\bea{\begin{eqnarray} }           \def\eea{\end{eqnarray} }
\def\beann{\begin{eqnarray*} }        \def\eeann{\end{eqnarray*} }
\def\beal{\begin{eqalign}}            \def\eeal{\end{eqalign}}
             
\def\bsubeq{\begin{subequations}}     \def\esubeq{\end{subequations}}
\def\bitem{\begin{itemize}}           \def\eitem{\end{itemize}}
                    
\def\pa{\partial}
\def\a{\alpha}
\def\b{\beta}

\def\d{\delta}
\def\g{\gamma}

\def\m{\mu}
\def\n{\nu}
\def\o{\omega}
\def\r{\rho}                    
\def\s{\sigma}                  

\def\G{\Gamma}

\def\L{\Lambda}

\usepackage{mathtools}
\usepackage{amsmath}

\def\in{\int d^{4}x\;e\;}
\def\nb{\nabla_{\beta}\psi}
\def\ns{\nabla_{\sigma}\psi}
\def\nea{(\nabla_{\alpha}e_{\mu}^{a})}
\def\neb{(\nabla_{\beta}e_{\nu}^{b})}

\def\mass{\left(m-\frac{2}{l}\right)}
\def\massl{\left(\frac{m}{l}-\frac{2}{l^{2}}\right)}
\def\massll{\left(\frac{m}{l^{2}}-\frac{2}{l^{3}}\right)}

\makeatletter
\@addtoreset{equation}{enumi}
\makeatother

\usepackage{slashed}
\usepackage{scalerel}
\usepackage{stackengine}
\stackMath
\newcommand\reallywidehat[1]{%
\savestack{\tmpbox}{\stretchto{%
  \scaleto{%
    \scalerel*[\widthof{\ensuremath{#1}}]{\kern-.6pt\bigwedge\kern-.6pt}%
    {\rule[-\textheight/2]{1ex}{\textheight}}
  }{\textheight}%
}{0.5ex}}%
\stackon[1pt]{#1}{\tmpbox}%
}
\begin{document}
\vspace{1.5cm}
\begin{titlepage}

\begin{center}

{\LARGE{\bf Dirac field and gravity in NC $SO(2,3)_\star$ model}}

\vspace*{1.3cm}

{{\bf Dragoljub Go\v canin  and
Voja Radovanovi\' c}}

\vspace*{1cm}

University of Belgrade, Faculty of Physics\\
Studentski trg 12, 11000 Beograd, Serbia \\[1em]

\end{center}

\vspace*{2cm}

\date{\today}

\begin{abstract}
Action for the Dirac spinor field coupled to gravity on noncommutative (NC) 
Moyal-Weyl space-time is obtained without prior knowledge of the metric tensor. 
We emphasise gauge origins of gravity (i.e. metric structure) and its interaction with fermions by demonstrating 
that a classical action invariant under $SO(2,3)$ gauge transformations can be 
exactly reduced to the Dirac action in curved space-time after breaking the original 
symmetry down to the local Lorentz $SO(1,3)$ symmetry. The commutative, $SO(2,3)$ invariant action 
can be straightforwardly deformed via Moyal-Weyl $\star$-product to its NC $SO(2,3)_\star$ invariant version which can be expanded perturbatively in the powers of the deformation parameter using the Seiberg-Witten map. 
The gravity-matter couplings in the expansion arise as an effect of the gauge symmetry breaking.  
We calculate in detail the first order NC correction to the classical Dirac action in
curved space-time and show that it does not vanish. This significant feature of the presented model enables us to potentially observe the NC effects already at the lowest perturbative order. Moreover, NC effects are apparent even in the flat space-time limit. We analyse NC modification of the Dirac equation, Feynman propagator and dispersion relation for electrons in Minkowski space-time. 
\end{abstract}
\vspace*{1cm}


\vspace*{1cm}
\quad\scriptsize{eMail:
dgocanin@ipb.ac.rs and rvoja@ipb.ac.rs}
\vfill

\end{titlepage}

\setcounter{page}{1}
\newcommand{\Section}[1]{\setcounter{equation}{0}\section{#1}}
\renewcommand{\theequation}{\arabic{section}.\arabic{equation}}

\section{Introduction}
\hspace{0.4cm} Quantum Field Theory (QFT) and General Relativity (GR) are two 
cornerstones of modern 
theoretical physics. Although these theories have been tested to an excellent 
degree of accuracy in 
their respective areas of applicability, occurrence of singularities in both of 
them strongly 
 indicates that they are incomplete. GR, as a classical theory of gravity, 
describes large-scale
 geometric structure of space-time and its relation to the distribution of 
matter. On the other hand,
 QFT, standing on the principles of Quantum Mechanics and Special Relativity, 
provides us with the
 Standard Model of elementary particles which successfully utilizes the idea of 
local symmetry to 
describe the fundamental particle interactions. Understanding quantum nature of 
space-time and uniting 
gravity with other fundamental interactions is considered to be one of the 
main goals of contemporary physics.\\[0.2cm]
In order to obtain a consistent unified theory, certain modifications of the 
basic concepts of QFT 
and GR are necessary. Various approaches have been proposed so far, stemming 
from String Theory, 
Loop Quantum Gravity, Noncommutative (NC) Field Theory, etc. and all of them,
 in some radical way, change the notion of point particle and/or that of space-time. \\[0.2cm]
In the last twenty years, Noncommutative Field Theory has become a very 
important direction of investigation in theoretical high energy physics and 
gravity. Its basic insight 
is that the quantum nature of space-time, at the microscopic level, should mean 
that even the 
space-time coordinates are to be treated as mutually incompatible observables, 
satisfying some non 
trivial commutation relations. The simplest choice of noncommutativity is the 
so 
called canonical noncommutativity, defined by  
\be [x^\m, x^\n]=i\theta^{\m\n}\ ,\label{can-kom-rel}\ee
where $\theta^{\m\n}$ are components of a constant antisymmetric matrix. \\[0.2cm]
To establish canonical noncommutativity, instead of using abstract 
algebra of coordinates, i.e. noncommutative space-time, one can equivalently introduce the 
noncommutative Moyal-Weyl $\star$-product,  
\begin{equation}
\label{moyal}  f (x)\star  g (x) =
      e^{\frac{i}{2}\,\theta^{\a\b}\frac{\pa}{\pa x^\a}\frac{\pa}{ \pa
      y^\b}} f (x) g (y)|_{y\to x}\ ,
\end{equation}
as a multiplication of functions (fields) defined on the usual, commutative 
space-time. 
The quantity $\theta^{\mu\nu}$ is considered to
be a small deformation parameter that has dimensions of $(\textit{length})^{2}$ (in natural units). It is a fundamental constant, like the Plank 
length or the speed of light. \\[0.2cm]
Recently, a lot of attention has been devoted to NC gravity, and many different approaches 
to this problem have been developed.   
In \cite{SWmapApproach} a deformation of pure Einstein gravity based on the 
Seiberg-Witten map is proposed.
Twist approach to noncommutative gravity was explored in 
\cite{TwistApproach,TwistSolutions}. Lorentz symmetry in NC field 
theories was considered in \cite {Chaichian}.
Some other proposals are given in 
\cite{EmGravityApproach,OtherApproaches,Faizal,Ostali,Dobrski,MajaJohn}. 
The connection to supergravity was established in \cite{PLSUGRA}. The extension of NC gauge theories to orthogonal and symplectic algebra was considered
 in \cite{ext-symplectic-bonora}.
Finally, in the previous papers of one of the authors  
\cite{UsLetter,Us-16,MiAdSGrav,MDVR-14,PLM-13} an
approach based on the deformed Anti-de Sitter (AdS) symmetry group, i.e. $SO(2,3)_{\star}$ group,  
and canonical noncommutativity was established. This approach emphasises gauge 
origins of gravity, which becomes manifest only after the suitable symmetry 
breaking. The action was constructed without previous introduction of the 
space-time metric structure and the second order NC correction to 
the Einstein-Hilbert action was found explicitly. Special attention has been 
devoted to the meaning of the coordinates used. 
Namely, it was shown that coordinates in which we postulate canonical
noncommutativity are the Fermi inertial coordinates, i.e. coordinates of an inertial observer along the geodesic. 
The commutator between arbitrary coordinates can be derived from the canonical 
noncommutativity as demonstrated in \cite{UsLetter}. \\[0.2cm]
A next natural step is to consider coupling of matter fields and gravity 
in the framework of NC $SO(2,3)_\star$ model. In this paper we specifically focus on the NC coupling of 
the Dirac spinor field and gravity.    
Previously, this problem has been treated by Aschieri and 
Castellani \cite{PLGR-fer}. Their model, based on the local 
$SO(1,3)_\star$ symmetry, is
significantly different from the one that we want to present here. One of the formal differences is that, in their approach, the vierbein field $e_{\m}^{a}$ is an adjoint field, i.e. it transforms in the adjoint representation of $SO(1,3)$ group, whereas in our approach, the vierbein and the $SO(1,3)$ spin-connection are just different components of the total $SO(2,3)$ gauge potential, and they are both being treated on equal footing. Therefore, in our model, the vierbein field holds the same status and transforms in the same way as an ordinary gauge potential. More importantly, physical 
implications to which these to models lead are quite different. The mentioned authors have found that the first non vanishing NC correction of the Dirac action in curved space-time is at the second order in powers of $\theta^{\a\b}$ (all odd-power corrections being equal to zero). The absence of the first order NC correction is a generic property of NC field theories that posses deformed Lorentz symmetry, and this makes them computationally challenging. Our model however entails non vanishing first order correction of the Dirac action in curved space-time that survives even in the flat space-time limit, and this leads us to an important physical prediction of the linearly deformed dispersion relation for electrons in Minkowski space-time and along with the Zeeman-like splitting of the classical energy levels in NC background. This result enables us to investigate potentially observable NC effects much more easily. As an aside, we should also mention that the differences between these two models revealed themselves already in the case of a pure gravity. Namely, we showed that the deformation of Minkowski space-time is obtained in NC $SO(2,3)_{\star}$ model \cite{UsLetter}. \\ [0.2cm]
The paper is organized as follows. In the next section we introduce the 
basic elements of AdS algebra and present a model of classical (undeformed) action based on 
the local AdS symmetry. 
In the third section we shortly review the theory of gauge fields on NC 
space-time. 
In section 4, we upgrade our classical action to its noncommutative (deformed) version. 
Since the noncommutativity is assumed to be small, we use the Seiberg-Witten (SW) map to expand 
the 
NC action perturbatively in powers of the deformation parameter $\theta^{\a\b}$, and calculate in detail the first order NC 
correction to the Dirac action in curved space-time. Finally, 
in section 5, we consider a special case of flat space-time and analyse the NC 
correction of the Dirac equation, Feynman propagator and dispersion relation for electrons. 
Section 6 contains 
discussion and conclusion.

\initiate 
\section{Commutative model}

\noindent
Before we introduce the model of commutative (i.e. classical) action based on the local $SO(2,3)$ 
symmetry, we will present some basic definitions concerning the algebra of $SO(2,3)$ 
group. Many more details can be found in our previous papers \cite{UsLetter,Us-16,MiAdSGrav,MDVR-14}.\\[0.2cm]
The generators of $SO(2,3)$ group are denoted by $M_{AB}$, where gauge group indices $A,B,...$ take values $0,1,2,3,5$. These generators satisfy the following commutation relations:
\be
[M_{AB},M_{CD}]=i(\eta_{AD}M_{BC}+\eta_{BC}M_{AD}-\eta_{AC}M_{BD}-\eta_{BD}M_{AC
})\ ,
\label{AdSalgebra}
\ee
where $\eta_{AB}={\rm diag}(+,-,-,-,+)$ is the $5D$ internal space metric 
tensor.
A realisation of this algebra can be obtained from $5D$ gamma matrices. 
Namely, if $\G^A$ are $5D$ gamma matrices that satisfy anticommutation 
relations: 
\be\{\G^A,\G^B\}=2\eta^{AB}\ ,\ee
then generators $M_{AB}$ are given by
\be 
M_{AB} =\frac{i}{4}[\Gamma_A,\Gamma_B]\ .
\ee 
One choice of $5D$ gamma matrices is $\Gamma_A =(i\gamma_a\gamma_5, \gamma_5)$, 
where $\g_a$ are the usual $4D$ gamma matrices. The local Lorentz indices $a,b,...$ take values $0,1,2,3$. 
In this particular representation, $SO(2,3)$ generators are given by
\bea
M_{ab} &=&\frac{i}{4}[\gamma_a,\gamma_b]=\frac12\sigma_{ab}\ ,\nonumber\\
M_{5a} &=&\frac{1}{2}\gamma_a\ . \label{Maba5}
\eea
The total $SO(2,3)$ gauge potential, $\omega_\m=\frac{1}{2}\omega_\m^{\;\;AB}M_{AB}$,
can be decomposed as  
\be\omega_\m=\frac{1}{4}\omega_\m^{\;\;ab}\sigma_{ab}-
\frac{1}{2l} e_\m^{a}\gamma_a\ ,\label{GaugePotAdsDecomp}
\ee
where $e_\mu^a$ and $\omega_\m^{\;\;ab}$ are the vierbein and the $SO(1,3)$ spin-connection, respectively, and $l$ is a constant length scale. The world indices 
$\m,\n,...$ take values $0,1,2,3$.
We see that along with the spin-connection, which is naturally related 
to the gauged $SO(1,3)$ group and suitable for introducing fermionic spinor fields in 
curved space-time, here we get additional gauge field - the vierbein 
which is related to the metric tensor by
\be
\eta_{ab}e_{\m}^{a}e_{\n}^{b}=g_{\m\n}\ ,\;\;\;\; e=\det(e_{\m}^{a})=\sqrt{-g}\ .
\ee
Thus, in this model, the vierbein and the $SO(1,3)$ spin-connection are just different components of the total $SO(2,3)$ gauge potential, and they are both being treated on equal footing. \\[0.2cm]
The field strength tensor is build from the gauge potential in the usual way:
\be
F_{\m\n}=\pa_\m\omega_\n-\pa_\n\omega_\m-i[\omega_\m,\omega_\n]
=\frac{1}{2}F_{\mu\nu}^{\;\;\;\;AB}M_{AB}\ . \label{FAB}
\ee
Its components are: 
\bea
F_{\m\n}^{\;\;\;\;ab}&=& R_{\m\n}^{\;\;\;\;
ab}-\frac{1}{l^2}(e_\m^ae_\n^b-e_\m^be_\n^a)\ ,\nn\\
F_{\m\n}^{\;\;\;\;a5}&=&\frac{1}{l}T_{\m\n}^{\;\;\;\;a}\ , \label{FabFa5}
\eea
where we recognize
\bea
R_{\m\n}^{\;\;\;\;ab} &=&
\pa_\m\omega_\n^{\;\;ab}-\pa_\n\omega_\m^{\;\;ab}+\omega_\m^{\;\;ac}\omega_\n^{\;\;cb}
-\omega_\m^{\;\;bc}\omega_\n^{\;\;ca} \ , \label{Rab}\\
T_{\m\n}^{\;\;\;\;a} &=& \nabla_\m e^a_\n-\nabla_\n e^a_\m\, \label{Ta}
\eea 
as the curvature tensor and the torsion, respectively. \\[0.2cm]
In the papers of Stelle, West and 
Wilczek \cite{Wilczek,stelle-west} a commutative action for pure gravity 
with $SO(2,3)$ 
gauge symmetry was constructed. Also, in the papers of Chamseddine and 
Mukhanov GR is formulated by gauging $SO(1,4)$ or, more suitable for 
supergravity,
$SO(2,3)$ group \cite{Mukhanov}. Proceeding within this general 
framework, which is motivated by the idea of constructing a unified symmetry setup for general relativity and gauge field theories, we show that it can also accommodate fermionic matter fields, specifically, 
the Dirac spinor field. We are going to do that by providing a classical (i.e. noncommutative) action for the Dirac spinors invariant under ordinary $SO(2,3)$ gauge 
transformations which exactly reduces to the standard Dirac action in curved space-time after the suitable symmetry breaking. Perturbative expansion via SW map of the deformed $SO(2,3)_{\star}$ invariant action will give us the NC corrections to the classical Dirac action in curved space-time after the symmetry breaking. \\[0.2cm]   
Let $\psi$ be a Dirac spinor field which transforms according to the fundamental 
representation of $SO(2,3)$ gauge group, i.e.
\be \d_{\epsilon} \psi=i\epsilon\psi=\frac{i}{2}\epsilon^{AB}M_{AB}\psi\ ,\ee
where $\epsilon^{AB}$ are antisymmetric gauge parameters.
The covariant derivative of a Dirac spinor is given by
\be D_{\m}\psi=\partial_{\m}\psi-\frac{i}{2}\o_{\m}^{\;\;AB}M_{AB}\psi \ .\ee 
Taking a hermitian conjugate of the previous expression we get 
\be 
D_{\m}\bar\psi=\partial_{\m}\bar\psi+\frac{i}{2}\bar\psi 
\;\o_{\m}^{\;\;AB}M_{AB}\ 
.
\ee
In order to break $SO(2,3)$ gauge symmetry \cite{stelle-west} we introduce an
auxiliary field 
$\phi=\phi^A\G_A$. This field is a space-time scalar and an internal-space vector. It transforms according to the adjoint 
representation 
of $SO(2,3)$, i.e.
\be \d_{\epsilon}\phi=i[\epsilon, \phi]\ ,\ee
and it is constrained by the condition $\phi^{A}\phi_{A}=l^{2}$. Note that this field has  mass dimension $-1$. The covariant derivative of an adjoint field is given by 
\begin{equation}
\label{cov-dev-adj} D_{\m}\phi=\partial_{\m}\phi-i[\o_\m,\phi]\ .
\end{equation} 
Consider the following kinetic-type symmetric-phase action: 
\be
  S_{kin}=\frac{i}{12}\int 
d^{4}x\;\varepsilon^{\mu\nu\rho\sigma}\Big[\bar{\psi}D_{\mu}\phi D_{\nu}\phi 
D_{\rho}\phi D_{\sigma}\psi
-D_{\sigma}\bar{\psi}D_{\mu}\phi D_{\nu}\phi D_{\rho}\phi\psi\Big]\ . 
\label{Classical-kinetic-action}\ \ee
This action is invariant under $SO(2,3)$ gauge transformations, and it is hermitian up to the surface term which vanishes. \\[0.2cm]
It is straightforward to show that the total covariant derivative of a spinor field can be 
separated like
\be D_\s\psi=\nabla_{\s}\psi+\frac{i}{2l}e_\s^{a}\gamma_{a}\psi \ ,\ee
where 
\be \nabla_{\s}\psi=\pa_\s\psi-\frac{i}{4}\o_\s^{\;\;ab}\s_{ab}\psi\ee
is the usual $SO(1,3)$ covariant derivative. \\[0.2cm]
We break the $SO(2,3)$ symmetry down to the local 
Lorentz $SO(1,3)$
symmetry by fixing the value of the auxiliary field $\phi$, specifically by taking $\phi^a=0$ and $\phi^5=l$. According to (\ref{cov-dev-adj}), 
the components of $D_\m\phi$ become 
$(D_\m\phi)^{a}=e_\m^a$ and $(D_\m\phi)^{5}=0$ and thus we obtain the action in the broken-symmetry phase:
\begin{equation} \label{Dirac in curved space}
S_{kin}=\frac{i}{2}\int d^{4}x\;e\;\left[\bar{\psi}\gamma^{\sigma}
\nabla_{\sigma}\psi-\nabla_{\sigma}\bar{\psi}\gamma^{\sigma}\psi\right]
-\frac{2}{l}\int d^{4}x\;e\;\bar{\psi}\psi \ ,
\end{equation}
which is exactly the Dirac action in curved space-time for 
spinors of mass  $2/l$\ . \\[0.2cm]
But, we know that leptons and quarks do not have the 
same 
masses. To obtain the correct masses of fermionic particles we have to 
include additional terms in the action. There are five terms invariant 
under $SO(2,3)$ transformations that can be used to supplement the original 
action in order to obtain the correct Dirac mass term in curved space-time after the symmetry 
breaking. These terms only differ in the position of the auxiliary filed $\phi$, and they are:
\bea &&\bar{\psi}D_{\mu}\phi D_{\nu}\phi 
D_{\rho}\phi D_{\sigma}\phi\phi\psi \ ,\ 
\bar{\psi}D_{\mu}\phi D_{\nu}\phi 
D_{\rho}\phi\phi D_{\sigma}\phi\psi \ ,\nn\\ 
&&\bar{\psi}D_{\mu}\phi D_{\nu}\phi 
\phi D_{\rho}\phi D_{\sigma}\phi\psi \ ,\ 
\bar{\psi}D_{\mu}\phi D_{\nu}\phi 
D_{\rho}\phi D_{\sigma}\phi\phi\psi \ ,\nn\\ 
&&\bar{\psi}D_{\mu}\phi\phi D_{\nu}\phi 
D_{\rho}\phi D_{\sigma}\phi\psi \ . \eea
We can combine them so to construct only three independent hermitian "mass terms" (terms of the type $\bar{\psi}...\psi$) denoted by $S_{m,i}$ $(i=1,2,3)$:
 \bea S_{m,1}&=&\frac{i}{2}c_{1}\Big(\frac{m}{l}
-\frac{2}{l^{2}}\Big)\int 
d^{4}x\;\varepsilon^{\mu\nu\rho\sigma}\Big[\bar{\psi}D_{\mu}\phi D_{\nu}\phi 
D_{\rho}\phi 
D_{\sigma}\phi\phi\psi+\bar{\psi}\phi D_{\mu}\phi D_{\nu}\phi D_{\rho}\phi 
D_{\sigma}\phi\psi\Big] \ , \nn \\
S_{m,2}&=&\frac{i}{2}c_{2}\Big
(\frac{m}{l}-\frac{2}{l^{2}}\Big)\int 
d^{4}x\;\varepsilon^{\mu\nu\rho\sigma}\Big[\bar{\psi}D_{\mu}\phi D_{\nu}\phi 
D_{\rho}\phi\phi D_{\sigma}\phi\psi+\bar{\psi}D_{\mu}\phi\phi D_{\nu}\phi 
D_{\rho}\phi D_{\sigma}\phi\psi\Big] \ , \nn \\
S_{m,3}&=&i\;c_{3}\Big(\frac{m}{l}-\frac{2}{l^{2}}\Big)\int 
d^{4}x\;\varepsilon^{\mu\nu\rho\sigma}\;\;\bar{\psi}D_{\mu}\phi D_{\nu}\phi
\phi D_{\rho}\phi D_{\sigma}\phi\psi \ . \label{com-mass-terms}
\eea
The undetermined dimensionless coefficients $c_1,c_2$ and $c_3$ are introduced for generality, and they will be fixed later. \\[0.2cm]  
\newpage
\noindent
After the symmetry breaking, the sum of the mass terms in (\ref{com-mass-terms}), denoted by $S_{m}$, 
reduces to
\be
 S_{m}=\sum_{i=1}^{3}S_{m,i}=24(c_{2}-c_{1}-c_{3})\left(m-\frac{2}{l}
\right)\int d^{4}x\;e\;\bar{\psi}\psi \ . \label{mass-afterSB-coefs}
\ee
If we want this to be equal to  the Dirac mass term, i.e. to have
\be
S_{m}=-\left(m-\frac{2}{l}\right)\int d^{4}x\;e\;\bar{\psi}\psi \ , 
\label{mass-afterSB}
\ee
then coefficients $c_{1}$, $c_{2}$, and $c_{3}$ must satisfy the following 
constraint:
\be 
 c_{2}-c_{1}-c_{3}=-\frac{1}{24} \ .\label{constraint}
\ee
Terms in (\ref{Dirac in curved space}) and (\ref{mass-afterSB}) that have a 
factor of 
the cosmological mass $2/l$ in them cancel each other out, and therefore, the 
total symmetric-phase  commutative action 
\be S=S_{kin}+S_{m}\ , \label{tot-com-action}\ee 
where $S_{kin}$ is given in (\ref{Classical-kinetic-action}) and $S_{m}$ is 
the sum of all three terms in (\ref{com-mass-terms}),
exactly reduces after the symmetry breaking to the Dirac action for spinors of mass $m$ in curved 
space-time,
\begin{equation}
S=\frac{i}{2}\int d^{4}x\;e\;\left[\bar{\psi}\gamma^{\sigma}
\nabla_{\sigma}\psi-\nabla_{\sigma}\bar{\psi}\gamma^{\sigma}\psi\right]
-m\int d^{4}x\;e\;\bar{\psi}\psi \ .
\end{equation}
Thus, by starting with a theory with $SO(2,3)$ gauge symmetry, by a suitable "gauge fixing", we have obtained the standard minimal coupling of the massive Dirac spinor field and gravity.

\initiate
\section{Gauge theories and the Seiberg-Witten map}

\hspace{0.4cm} Let us briefly review the theory of gauge fields on noncommutative space-time by summarising the most relevant results. Our approach is based on the use of the Seiberg-Witten map \cite{SWMapEnvAlgebra}.   
In analogy to the ordinary (commutative) gauge field theory, one introduces NC spinor field $\widehat{\psi}$, NC adjoint field $\widehat{\phi}$ and NC gauge potential $\widehat{\o}_\m$. We use NC gauge potential to construct NC field strength, 
\be \widehat{F}_{\m\n}=\pa_\m\widehat{\omega}_\n-\pa_\n\widehat{\omega}_\m
-i[\widehat{\omega}_\m\ds\widehat{\omega}_\n] \ . 
\ee\label{nckrivina}
The covariant derivatives of NC adjoint and spinor field are defined by 
\bea
D_\m\widehat{\phi}&=&\pa_\m \widehat{\phi}-i[\widehat{\omega}_\m\ds 
\widehat{\phi}] \ , \\
D_\m\widehat{\psi}&=&\pa_\m \widehat{\psi}-i\widehat{\omega}_\m 
\star\widehat{\psi} \ .
\eea 
Note that the structure of covariant derivatives for both fields is the same as in the commutative field theory, the only difference being the use of the Moyal-Weyl $\star$-product instead of the ordinary multiplication. \\[0.2cm]    
Under infinitesimal gauge transformations the noncommutative fields 
$\widehat\psi$ and $\widehat{\phi}$, along with their covariant derivatives, transform according to the fundamental and adjoint representation, respectively, i.e.  
\begin{align}
\delta^\star_\epsilon \widehat {\psi}&= i \widehat{\Lambda}_\epsilon 
\star{\widehat
\psi} \ ,\;\;\;\;\delta^\star_\epsilon D_\m \widehat{\psi}=i\widehat{\L}_\epsilon\star 
D_\m\widehat{\psi} \ , \nn\\
\delta^\star_\epsilon {\widehat \phi}&= i[\widehat{\Lambda}_\epsilon 
\ds{\widehat
\phi}] \ ,\;\;\;
\delta^\star_\epsilon D_\m\widehat{\phi}=i[\widehat{\L}_\epsilon\ds 
D_\m\widehat{\phi}] \ . 
\label{inf-transf}\end{align}
The transformation laws for NC gauge potential and NC field strength are given by 
\bea
\delta^\star_\epsilon \widehat {\o}_{\m}&=& 
\pa_\m\widehat{\L}_\epsilon-i[\widehat{\o}_\m\ds\widehat{\L}_\epsilon] \ , \\ 
\delta^\star_\epsilon \widehat {F}_{\m\n}&=& i[\widehat{\Lambda}_\epsilon 
\ds \widehat
{F}_{\m\n}] \ .
\eea
We see that NC field strength $\widehat{F}_{\m\n}$ transforms according to the adjoint representation of $SO(2,3)_{\star}$ just as ordinary field strength $F_{\m\n}$ transforms according to the adjoint representation of $SO(2,3)$. In the previous transformation rules, 
$\widehat{\Lambda}_\epsilon$ is a NC gauge parameter, 
and $\epsilon$ a commutative gauge parameter. \\ [0.2cm]
There is a relation between noncommutative and commutative fields 
known as the 
Seiberg-Witten map. It is based on the relation between commutative 
and noncommutative symmetries. The noncommutative quantities can be 
represented as power series in the deformation parameter $\theta^{\m\n}$, with expansion coefficients built out of the commutative quantities: 
$\epsilon, \phi,\psi$ and $\omega_\m$.   
\bea
{\widehat\omega}_\m &=& \omega_\mu
-\frac{1}{4}\theta^{\a\b}\{\omega_\a, \partial_\b\omega_\mu +
F_{\b\m}\}
+ {\cal O}(\theta^2)\ , \label{NC-Omega}  \\
{\widehat \phi} &=& \phi -\frac{1}{4}\theta^{\a\b}
\{\omega_\a,(\pa_\b + D_\b) \phi\} + {\cal O}(\theta^2) \ , \\
\widehat{\psi}&=&
\psi-\frac14\theta^{\a\b}\o_\a(\pa_\b+D_\b)\psi+{\cal 
O}(\theta^2) \ , \label{psi-expansion}\\
\widehat{\bar{\psi}}&=&\bar{\psi}-\frac14\theta^{\alpha\beta}(\partial_{\beta}+D_{\beta})\bar{\psi}\omega_{\alpha}
+ {\cal O}(\theta^2) \ , \label{psibar-expansion}\\ 
\widehat{\L}_\epsilon&=&\epsilon
-\frac14\theta^{\a\b}\{\o_\a,\pa_\b\epsilon\}
+{
\cal O}(\theta^2)\ .
\eea
Using the SW map, we can derive the first order 
NC corrections to the field strength, and the covariant 
derivatives of adjoint and spinor field. They are given by
\begin{equation}\label{F expansion}
\widehat{F}_{\m\n}=F_{\m\n}-\frac{1}{4}\theta^{\alpha\beta}\{\omega_{\alpha} ,
(\partial_{\beta}+D_{\beta})F_{\m\n}\}
+\frac{1}{2}\theta^{\alpha\beta}\{F_{\alpha\mu},F_{\beta\nu}\}+{\cal 
O}(\theta^2)\ ,
\end{equation}
\begin{equation}\label{Phi expansion}
D_{\mu}\widehat{\phi}=D_\m\phi-\frac{1}{4}\theta^{\alpha\beta}\{\omega_{\alpha} 
,
(\partial_{\beta}+D_{\beta})
D_{\mu}\phi\}+\frac{1}{2}\theta^{\alpha\beta}\{F_{\alpha\mu},D_{\beta}\phi\}+{
\cal O}(\theta^2) \ ,
\end{equation}
\be\label{Psi expansion}
D_\m\widehat{\psi}=D_\m\psi-
\frac14\theta^{\a\b}\o_\a(\pa_\b 
+D_\b)D_\m\psi+\frac{1}{2}\theta^{\a\b}F_{\a\m}D_\b\psi+{\cal O}(\theta^2) \ .
\ee
All these results will be put into use in the next section where we turn to the perturbative expansion of NC action for the Dirac spinor field.
\newpage
\initiate
\section{NC action}
Now, we are going to deform the commutative symmetric-phase action (\ref{tot-com-action}) 
by
replacing ordinary commutative fields, $\phi$ and $\psi$, 
with their noncommutative counterparts,
 $\widehat{\phi}$ and $\widehat{\psi}$, and by applying the Moyal-Weyl star product 
defined in (\ref{moyal}) instead of the usual 
commutative multiplication. This NC action
 can be expanded perturbatively in 
 powers of the deformation parameter, assuming that this parameter is small. We will investigate the first order NC correction to the kinetic and 
the mass terms separately. It will
be demonstrated by explicit calculation that the first order NC correction after the symmetry breaking does not
vanish.

\subsection{NC deformation of the kinetic term}

The noncommutative version of the kinetic action (\ref{Classical-kinetic-action}) 
will be denoted by a "hat" symbol and it is given by
\begin{align}\label{NC-kinetic-action}
  \widehat{S}_{kin}=\frac{i}{12}\int 
d^{4}x\;\varepsilon^{\mu\nu\rho\sigma}\;\Big[\widehat{\bar{\psi}}
\star(D_{\mu}\widehat{\phi})&\star(D_{\nu}\widehat{\phi}
)\star(D_{\rho}\widehat{\phi})\star(D_{\sigma}\widehat{\psi}
) \nn\\
-(D_{\sigma}\widehat{\bar{\psi}})&\star(D_
{ \mu } \widehat{\phi} )\star 
(D_{\nu}\widehat{\phi})\star(D_{\rho}\widehat{\phi}
)\star\widehat{\psi}\Big] \ . 
\end{align} 
Using the infinitesimal transformation rules (\ref{inf-transf})
one can easily check that the action (\ref{NC-kinetic-action}) is invariant 
under deformed $SO(2,3)_\star$ gauge transformations. Moreover, 
this action is  hermitian up to the surface term which vanishes. \\[0.2cm]
Let us now expand the action (\ref{NC-kinetic-action}) 
up to the first order 
in the deformation parameter $\theta^{\a\b}$ using the 
Seiberg-Witten map. Generally, for any two NC fields
 $\widehat{A}$ and $\widehat{B}$, 
the first order NC correction of their product is given by
\begin{equation} 
\left(\widehat{A}\star\widehat{B}\right)^{(1)}
=\widehat{A}^{(1)}B+A\widehat{B}^{(1)}
+\frac{i}{2}\theta^{\alpha\beta}\partial_\a A \partial_\b B \ . \label{rule}
\end{equation}
If both of these two fields transform according 
to the adjoint representation, the last formula takes on the specific form,
\begin{align}
 \left(\widehat{A}\star\widehat{B}\right)^{(1)}=
 &-\frac{1}{4}\theta^{\a\b}\{\omega_\a,(\partial_\b+D_\b)AB\}
 +\frac{i}{2}\theta^{\a\b} D_\a A D_\b B \nn \\
 &+cov(\widehat{A}^{(1)})B+Acov(\widehat{B}^{(1)}) \ , \label{rule1}
\end{align}    
where $cov(\widehat{A}^{(1)})$ is the covariant part of $A's$ first order NC correction, and $cov(\widehat{B}^{(1)})$, 
the covariant part of $B's$ first order NC correction. 
Applying the rule (\ref{rule1}) twice, and using the 
expansion $(\ref{Phi expansion})$ for the covariant 
derivative of the adjoint field $\phi$, 
we can obtain the first order NC correction of the product $D_{\mu}\widehat{\phi}\star
D_{\nu}\widehat{\phi}\star D_{\rho}\widehat{\phi}$\ :
\begin{align} \label{triple-product}
\left(D_{\mu}\widehat{\phi}\star
D_{\nu}\widehat{\phi}\star D_{\rho}\widehat{\phi}\right)^{(1)}
=&-\frac{1}{4}\theta^{\alpha\beta}\{\omega_{\alpha},(\partial_{\beta}
+D_{\beta})(D_{\mu}\phi
D_{\nu}\phi D_{\rho}\phi)\} \nn \\
&+\frac{i}{2}\theta^{\alpha\beta}D_{\alpha}(D_{\mu}\phi
D_\nu\phi)(D_{\beta}D_{\rho}\phi)+\frac{i}{2}\theta^{\alpha\beta}(D_{\alpha}D_{\mu}\phi)(D_{\beta}
D_{\nu}\phi)D_{\rho}\phi \nn \\ 
&+\frac{1}{2}\theta^{\alpha\beta}\{F_{\alpha\mu},D_{\beta}\phi\}D_{\nu}\phi D_{\rho}\phi+\frac{1}{2}\theta^{\alpha\beta}D_{\mu}\phi
\{F_{\alpha\nu},D_{\beta}\phi\}
D_{\rho}\phi \nn \\
&+\frac{1}{2}\theta^{\alpha\beta}D_{\mu}\phi 
D_{\nu}\phi\{F_{\alpha\rho},D_{\beta}\phi\} \ .
\end{align} 
Note that composite field $D_{\mu}\widehat{\phi}\star
D_{\nu}\widehat{\phi}\star D_{\rho}\widehat{\phi}$ 
also transforms according to the adjoint representation of $SO(2,3)_{\star}$ 
since it is a product of the fields that transform according to the adjoint representation.
Thus, according to the rule (\ref{rule1}), we could immediately say what is the non-covariant part in the first order NC correction to $D_{\mu}\widehat{\phi}\star
D_{\nu}\widehat{\phi}\star D_{\rho}\widehat{\phi}$, i.e. what is the first term in (\ref{triple-product}). It is non-covariant because of the way in which it incorporates the gauge potential $\omega_{\a}$ and the  partial derivative $\partial_{\b}$. The other terms appearing in (\ref{triple-product}) 
are manifestly covariant. \\[0.2cm]
If we have a field $\widehat{A}$ that transforms according to the adjoint representation,
and field $\widehat{B}$ that transforms according to the 
fundamental representation, then rule (\ref{rule}) again takes on the specific form,
\begin{align}
 \left(\widehat{A}\star\widehat{B}\right)^{(1)}=&-\frac{1}{4}\theta^{\a\b}\omega_\a(\partial_\b+D_\b)(AB)
 +\frac{i}{2}\theta^{\a\b} D_\a A D_\b B \nn \\
 &+cov(\widehat{A}^{(1)})B+Acov(\widehat{B}^{(1)}) \ . \label{rule-fund-rep}
\end{align} 
Using the result (\ref{triple-product}) and the 
expansion $(\ref{Psi expansion})$ for the covariant 
derivative of a spinor field, we 
can obtain the first order NC correction of the noncommutative product $
D_{\m}\widehat{\phi}\star 
D_{\n}\widehat{\phi}\star D_\r\widehat{\phi}\star 
D_\r\widehat{\psi}$. Applying the rule (\ref{rule-fund-rep}), and setting $\widehat{A}:=
D_{\m}\widehat{\phi}\star
D_{\n}\widehat{\phi}\star D_\r\widehat{\phi}$ and $\widehat{B}:=D_\s\widehat{\psi}$, we get  
\begin{eqnarray}
(D_{\m}\widehat{\phi}\star
D_{\n}\widehat{\phi}\star D_\r\widehat{\phi}\star
D_\s\widehat{\psi})^{(1)}=&-&\frac{1}{4}\theta^{\a
\beta}\omega_{\alpha}(\partial_{\beta}+D_{\beta})(D_{\m}\phi
D_{\n} D_\r\phi D_\s\psi) \nn \\
&+&\frac{i}{2}\theta^{\alpha\beta}D_{\alpha}(D_{\m}\phi D_\n\phi D_\r\phi)(D_{\beta}D_{\s}\psi) \nn \\
&+&\frac{i}{2}\theta^{\a\b}D_\a( D_\m \phi  D_\n\phi)
 (D_\b D_\r\phi) D_\s\psi\nn\\ 
 &+&\frac{i}{2}\theta^{\a\b}(D_\a D_\m \phi)
 (D_\b D_\n\phi) D_\r\phi D_\s\psi \nn\\
&+&\frac{1}{2}\theta^{\alpha \beta} 
\{F_{\alpha\m},D_{\beta}\phi\}D_\n \phi D_{\r}\phi D_\s\psi \nn \\
&+&\frac{1}{2}\theta^{\alpha\beta}D_{\m}\phi  
\{F_{\alpha \n},D_{\beta}\phi\}D_\r\phi D_\s\psi \nn\\
&+&\frac{1}{2}\theta^{\alpha
\beta}D_{\m}\phi D_\n\phi \{F_{\alpha \r},D_{\beta}\phi\}D_\s\psi \nn \\
&-&\frac{1}{2}\theta^{\alpha\beta}D_{\m}\phi D_\n\phi D_\r\phi F_{\s\alpha
} D_{\beta}\psi \ . \label{pp}
\end{eqnarray}
The composite field $D_\m\widehat{\phi}\star D_\n\widehat{\phi}\star D_\r\widehat{\phi}\star 
D_\s\widehat{\psi}$ transforms according to the spinor representation 
since it is a product of the field $D_\m\widehat{\phi}
\star D_\n\widehat{\phi}\star D_\r\widehat{\phi}$ that transforms 
according to the adjoint 
representation, and the field $D_\s\widehat{\psi}$ that transforms
 according to the fundamental representation, and for that reason 
the first term 
in (\ref{pp}), i.e. the non-covariant term,     
has the same form as the corresponding non-covariant term in (\ref{Psi expansion}). Again, we knew that from the general result (\ref{rule-fund-rep}). The other terms in (\ref{pp}) are manifestly covariant. \\[0.2cm]
\newpage
\noindent
At last, by using the NC expansion of the Dirac adjoint field $(\ref{psibar-expansion})$, setting $\widehat{A}:=\widehat{\bar{\psi}}$ and $\widehat{B}:=D_\m\widehat{\phi}\star
D_\n\widehat{\phi}\star D_\r\widehat{\phi}\star D_\s\widehat{\psi}$, 
the general rule (\ref{rule}) gives us the total product $\widehat{\bar{\psi}}\star D_\m\widehat{\phi}\star
D_\n\widehat{\phi}\star D_\r\widehat{\phi}\star D_\s\widehat{\psi}$ which is a scalar of the deformed $SO(2,3)_{\star}$ gauge group:
\begin{eqnarray}
\widehat{\bar{\psi}}\star D_\m\widehat{\phi}\star
D_\n\widehat{\phi}\star D_\r\widehat{\phi}\star D_\s\widehat{\psi}=&-&\frac{1}{4}\theta^{\alpha\beta}\bar\psi 
F_{\alpha\beta}D_{\m}\phi
D_{\n} D_\r\phi D_\s\psi\nn\\
&+&\frac{i}{2}\theta^{\alpha\beta}\bar\psi D_{\alpha}(D_{\m}\phi
D_\n\phi D_\r\phi)(D_{\beta}D_{\s}\psi)\nn\\&+&
\frac{i}{2}\theta^{\alpha\beta}\bar\psi D_\a( D_\m \phi D_\n
\phi) (D_\b D_\r\phi) D_\s\psi\nn\\
&+&\frac{i}{2}\theta^{\alpha\beta}\bar\psi(D_\a D_\m \phi)( D_\b
D_\n
\phi) D_\r\phi D_\s\psi\nn\\
&+&\frac{1}{2}\theta^{\alpha\beta}\bar\psi \{F_{\alpha
\m},D_{\beta}\phi\} D_\n\phi D_{\r}\phi
D_\s\psi\nn\\
&+&\frac{1}{2}\theta^{\alpha\beta}\bar\psi D_{\m}\phi  \{F_{\alpha \n},D_{\beta}\phi\} 
D_\r\phi
D_\s\psi\nn\\
&+&
\frac{1}{2}\theta^{\alpha\beta}\bar\psi D_\m\phi D_\n\phi  \{F_{\alpha
\r},D_{\beta}\phi\}
D_\s\psi \nn \\
&-&\frac{1}{2}\theta^{\alpha\beta}\bar\psi D_{\m}\phi D_\n\phi D_\r\phi
F_{\s\alpha
} D_{\beta}\psi\ .
\end{eqnarray} 

Finally, we present the first order NC correction of the commutative kinetic action in the symmetric phase, i.e. a $n=1$ term in the perturbative expansion of the full NC kinetic action $\widehat{S}_{kin}=\sum_{n}\widehat{S}_{kin}^{(n)}$:
\begin{eqnarray}
\widehat{S}^{(1)}_{kin}=\frac{i}{12}\;\theta^{\alpha\beta}\int 
d^{4}x\;\varepsilon^{\mu\nu\rho\sigma}\;\Bigg[&-&\frac{1}{4}\bar\psi 
F_{\alpha\beta}D_{\m}\phi
D_{\n} D_\r\phi D_\s\psi\nn\\
&+&\frac{i}{2}\bar\psi D_{\alpha}(D_{\m}\phi
D_\n\phi D_\r\phi)(D_{\beta}D_{\s}\psi)\nn\\&+&
\frac{i}{2}\bar\psi D_\a( D_\m \phi D_\n
\phi) (D_\b D_\r\phi) D_\s\psi\nn\\
&+&\frac{i}{2}\bar\psi(D_\a D_\m \phi)( D_\b
D_\n
\phi) D_\r\phi D_\s\psi\nn\\
&+&\frac{1}{2}\bar\psi \{F_{\alpha
\m},D_{\beta}\phi\} D_\n\phi D_{\r}\phi
D_\s\psi\nn\\
&+&\frac{1}{2}\bar\psi D_{\m}\phi  \{F_{\alpha \n},D_{\beta}\phi\} 
D_\r\phi
D_\s\psi\nn\\
&+&
\frac{1}{2}\bar\psi D_\m\phi D_\n\phi  \{F_{\alpha
\r},D_{\beta}\phi\}
D_\s\psi \nn \\
&-&\frac{1}{2}\bar\psi D_{\m}\phi D_\n\phi D_\r\phi
F_{\s\alpha
} D_{\beta}\psi\;\;\;\Bigg]+h.c. \ . \label{NCaction-beforeSB}
\end{eqnarray} 

This action possesses ordinary $SO(2,3)$, i.e. AdS symmetry, and this was to be expected by the virtue of the SW map. Namely, we started with the NC action (\ref{NC-kinetic-action}) invariant under the deformed $SO(2,3)_{\star}$ gauge transformations and expanded it perturbatively in 
powers of the deformation parameter $\theta^{\a\b}$ (up to the first order, but we could straightforwardly proceed further). By using the SW map we ensure that the obtained perturbative corrections are invariant, in each order, under the ordinary commutative $SO(2,3)$ gauge transformations. Our explicit result (\ref{NCaction-beforeSB}) confirms that.
\newpage 
\noindent
Setting $\phi^a=0$ and $\phi^5=l$, the symmetry of the action (\ref{NCaction-beforeSB}) is broken down to the local Lorentz $SO(1,3)$ symmetry and the action reduces to  
\bea\label{kinetic-term-afterSB}
  \widehat{S}^{(1)}_{kin}=\theta^{\a\b}\Bigg[&-&\frac{1}{8}\in R_{\alpha\mu}^{\;\;\;\;ab}e^{\mu}_{a}\;\bar{\psi}\gamma_{b}
   \nb +\frac{1}{16}\in R_{\alpha\beta}^{\;\;\;\;ab}e^{\sigma}_{b}\;\bar{\psi}\gamma_{a}\ns\nn\\
   &-&\frac{i}{32}\in R_{\alpha\beta}^{\;\;\;\;ab}
   \varepsilon_{abc}^{\;\;\;\;\;\;d}e^{\sigma}_{d}\;\bar{\psi}\gamma^{c}\gamma^{5}\ns
   -\frac{i}{16}\in R_{\alpha\mu}^{\;\;\;\;bc}e^{\mu}_{a}\varepsilon^{a}_{\;\;bcm}\;\bar{\psi}\gamma^{m}\gamma^{5}\nb \nn\\
   &-&\frac{i}{24}\in R_{\alpha\mu}^{\;\;\;\;ab}
   \varepsilon_{abc}^{\;\;\;\;\;\;d}e^{c}_{\beta}
   (e^{\mu}_{d}e^{\sigma}_{s}-e^{\mu}_{s}e^{\sigma}_{d})\;\bar{\psi}\gamma^{s}\gamma^{5}\ns\nn\\
   &-&\frac{i}{8l}\in T_{\alpha\beta}^{\;\;\;\;a}e^{\sigma}_{a}\;\bar{\psi}\ns
   +\frac{i}{8l}\in T_{\alpha\mu}^{\;\;\;\;a}e^{\mu}_{a}\;\bar{\psi}\nb\nn\\
   &+&\frac{1}{16l}\in T_{\alpha\beta}^{\;\;\;\;a}e^{\mu}_{a}\;\bar{\psi}\sigma_{\mu}^{\;\;\sigma}\ns
   +\frac{1}{8l}\in T_{\alpha\mu}^{\;\;\;\;a}e^{\mu}_{b}\;\bar{\psi}\sigma_{a}^{\;\;b}
   \nb\nn\\
   &-&\frac{1}{12l}\in T_{\alpha\mu}^{\;\;\;\;a}\varepsilon_{ab}^{\;\;\;\;cd}e_{\beta}^{b}e^{\mu}_{c}e^{\sigma}_{d}\;\bar{\psi}\gamma^{5}\ns
   +\frac{7i}{48l^{2}}\in\varepsilon_{abc}^{\;\;\;\;\;\;d}e^{a}_{\alpha}e^{b}_{\beta}e^{\sigma}_{d}\;\bar{\psi}\gamma^{c}\gamma^{5}\ns\nn\\
   &-&\frac{1}{4}\in \nea(e^{\mu}_{a}e^{\sigma}_{b}-e^{\sigma}_{a}e^{\mu}_{b})\;\bar{\psi}\gamma^{b}\nabla_{\beta}\nabla_{\sigma}\psi
   -\frac{1}{4l}\in \bar{\psi}\sigma_{\alpha}^{\;\;\sigma}\nabla_{\beta}
   \nabla_{\sigma}\psi\nn\\
   &-&\frac{i}{8}\in \eta_{ab}\nea\neb\varepsilon^{cdrs}e^{\mu}_{c}e^{\nu}_{d}e^{\sigma}_{s}\;\bar{\psi}\gamma_{r}
   \gamma_{5}\ns\nn\\
  &+&\frac{i}{12}\in(\nabla_{\alpha}e_{\mu}^{a})\neb\varepsilon_{b}^{\;\;cds}e^{\mu}_{c}e^{\nu}_{d}e^{\sigma}_{s}\;\bar{\psi}\gamma_{a}\gamma_{5}\ns\nn\\
   &-&\frac{1}{12l}\in e_{\alpha}^{c}(\nabla_{\beta}e^{b}_{\nu})
   \varepsilon_{bc}^{\;\;\;\;ds}e^{\nu}_{d}e^{\sigma}_{s}\;\bar{\psi}\gamma_{5}
   \ns \nn\\
   &-&\frac{1}{8l}\in\nea(e^{\mu}_{a}e^{\sigma}_{b}-e^{\sigma}_{a}e^{\mu}_{b})e^{c}_{\beta}\;\bar{\psi}\sigma^{b}_{\;\;c}\ns \nn\\
   &-&\frac{i}{2l}\in\nea e_{a}^{\mu}\;\bar{\psi}\nb
   -\frac{1}{8l}\in\nea e^{\mu}_{b}\;\bar{\psi}\sigma_{a}^{\;\;b}\nb 
\nn\\
   &+&\frac{1}{96l}\in R_{\alpha\beta}^{\;\;\;\;ab}\;\bar{\psi}\sigma_{ab}
   \psi\nn\\&-&\frac{5}{48l}\in R_{\alpha\mu}^{\;\;\;\;ab}e^{\mu}_{a}e^{c}_{\beta}\;\bar{\psi}\sigma_{bc}\psi
   -\frac{1}{16l}\in R_{\alpha\mu}^{\;\;\;\;ab}e_{\beta a}e^{\mu}_{c}\;\bar{\psi}\sigma_{b}^{\;\;c}\psi\nn\\
   &-&\frac{3}{32l^{2}}\in T_{\alpha\beta}^{\;\;\;\;a}\;\bar{\psi}\gamma_{a}\psi
   -\frac{1}{16l^{2}}\in T_{\alpha\mu}^{\;\;\;\;a}e^{\mu}_{a}\;\bar{\psi}\gamma_{\beta}\psi\nn\\
   &+&\frac{1}{16l^{2}}\in T_{\alpha\mu}^{\;\;\;\;a}e_{\beta a}\;\bar{\psi}\gamma^{\mu}\psi
      +\frac{1}{12l}\in \eta_{ab}\nea\neb\;\bar{\psi}\sigma^{\mu\nu}\psi\nn\\
         &-&\frac{1}{6l}\in \nea\neb (e^{\mu}_{a}e^{\nu}_{c}-e^{\mu}_{c}e^{\nu}_{a})\;\bar{\psi}\sigma^{c}_{\;\;b}\psi\nn\\
   &-&\frac{3}{16l^{2}}\in \nea e^{\mu}_{a}\;\bar{\psi}\gamma_{\beta}\psi
   +\frac{1}{16l^{2}}\in \nea e_{\beta a}\;\bar{\psi}\gamma^{\mu}\psi \nn\\
   &-&\frac{1}{3l^{3}}\in \bar{\psi}\sigma_{\alpha\beta}\psi\;\;\;\Bigg] + h.c. 
\ .
\eea
\newpage
\subsection{NC deformation of the mass terms}

\hspace{0.4cm} In this section we consider a noncommutative deformation 
of the mass terms. It 
is obtained by replacing the ordinary commutative product with 
the NC Moyal-Weyl $\star$-product in mass terms (\ref{com-mass-terms}):
\begin{align}
   \widehat{S}_{m}=\frac{i}{2}\Big(\frac{m}{l}
-\frac{2}{l^{2}}\Big)\int d^{4}x\;\varepsilon^{\mu\nu\rho\sigma}\bigg[\;&c_{1}\widehat{\bar{\psi}}\star D_{\mu}\widehat{\phi}\star D_{\nu}\widehat{\phi} \star D_{\rho}
\widehat{\phi}\star D_{\sigma}\widehat{\phi}\star\widehat{\phi}
\star\widehat{\psi} \nn\\
   +&c_{2}\widehat{\bar{\psi}}\star D_{\mu}\widehat{\phi}\star D_{\nu}\widehat{\phi}\star D_{\rho}\widehat{\phi}\star\widehat{\phi}\star D_{\sigma}\widehat{\phi}\star\widehat{\psi}\nn \\
+&c_{3}\widehat{\bar{\psi}}\star D_{\mu}\widehat{\phi}\star 
D_{\nu}\widehat{\phi}\star\widehat{\phi}\star D_{\rho}\widehat{\phi}\star 
D_{\sigma}\widehat{\phi}\star\widehat{\psi}\;\bigg]+h.c. \ .\label{NC mass term}
\end{align}
Again, by using the Seiberg-Witten map we can represent this action as a 
perturbation series in powers of the deformation parameter $\theta^{\alpha\beta}$, taking only first order term into account. We present the result for each of the three terms, denoted by $\widehat{S}^{(1)}_{m,i}$ $(i=1,2,3)$, separately:   
\begin{align}
 \widehat{S}^{(1)}_{m,1}=\frac{ic_{1}}{2}\Big(\frac{m}{l}
-\frac{2}{l^{2}}\Big)\;\theta^{\alpha\beta}\;\int 
d^{4}x\;\varepsilon^{\mu\nu\rho\sigma} 
\Bigg[\;
&+\frac{i}{2}\bar{\psi}D_{\alpha}(D_{\mu}\phi D_{\nu}\phi D_{\rho}\phi 
D_{\sigma}\phi\phi)D_{\beta}\psi \nn\\
&-\frac{1}{4}\bar{\psi}F_{\alpha\beta}
D_{\mu}\phi D_{\nu}\phi D_{\rho}\phi D_{\sigma}\phi\phi\psi \nn \\
&+\frac{i}{2}\bar{\psi}D_{\alpha}(D_{\mu}\phi D_{\nu}\phi D_{\rho}\phi D_{\sigma}\phi)D_{\beta}\phi\psi \nn\\
&+\frac{i}{2}\bar{\psi}D_{\alpha}(D_{\mu}\phi D_{\nu}\phi)D_{\beta}(D_{\rho}\phi D_{\sigma}\phi)\phi\psi \nn \\
&+\frac{i}{2}\bar{\psi}D_{\mu}\phi D_{\nu}\phi(D_{\alpha}D_{\rho}\phi)(D_{\beta}D_{\sigma}\phi)\phi\psi \nn\\
&+\frac{i}{2}\bar{\psi}(D_{\alpha}D_{\mu}\phi)(D_{\beta}D_{\nu}\phi)D_{\rho}\phi D_{\sigma}\phi\phi\psi \nn \\
&+\frac{1}{2}\bar{\psi}\{F_{\alpha\mu},D_{\beta}\phi\}D_{\nu}\phi 
   D_{\rho}\phi D_{\sigma}\phi\phi\psi \nn\\
&+\frac{1}{2}\bar{\psi}D_{\mu}\phi\{F_{\alpha\nu},D_{\beta}\phi\}D_{\rho}\phi D_{\sigma}\phi\phi\psi \nn \\
&+\frac{1}{2}\bar{\psi}D_{\mu}\phi D_{\nu}\phi\{F_{\alpha\rho},D_{\beta}\phi\}D_{\sigma}\phi\phi\psi \nn\\
&+\frac{1}{2}\bar{\psi}D_{\mu}\phi D_{\nu}\phi 
D_{\rho}\phi\{F_{\alpha\sigma},D_{\beta}\phi\}\phi\psi\;\Bigg] \ .\label{mass1}
\end{align}
\begin{align}
   \widehat{S}^{(1)}_{m,2}=\frac{ic_{2}}{2}\Big(\frac{m}{l}
-\frac{2}{l^{2}}\Big)\;\theta^{\alpha\beta}\;\int 
d^{4}x\;
\varepsilon^{\mu\nu\rho\sigma} 
\Bigg[&+\frac{i}{2}\bar{\psi}D_{\alpha}(D_{\mu}
\phi D_{\nu}\phi D_{\rho}\phi\phi D_{\sigma}\phi)D_{\beta}\psi \nn\\
   &-\frac{1}{4}\bar{\psi}F_{\alpha\beta}D_{\mu}\phi D_{\nu}\phi 
   D_{\rho}\phi\phi D_{\sigma}\phi\psi\nn \\
  &+\frac{i}{2}\bar{\psi}D_{\alpha}(D_{\mu}\phi D_{\nu}\phi D_{\rho}\phi\phi)(D_{\beta}D_{\sigma}\phi)\psi \nn\\
   &+\frac{i}{2}\bar{\psi}D_{\alpha}(D_{\mu}\phi D_{\nu}\phi D_{\rho}\phi)
   D_{\beta}\phi D_{\sigma}\phi\psi \nn\\
  &+\frac{i}{2}\bar{\psi}D_{\alpha}(D_{\mu}\phi D_{\nu}\phi)(D_{\beta}D_{\rho}\phi)\phi D_{\sigma}\phi\psi \nn\\
   &+\frac{i}{2}\bar{\psi}(D_{\alpha}D_{\mu}\phi)(D_{\beta}D_{\nu}\phi)
   D_{\rho}\phi\phi D_{\sigma}\phi\psi \nn\\
   &+\frac{1}{2}\bar{\psi}\{F_{\alpha\mu},D_{\beta}\phi\}D_{\nu}\phi 
   D_{\rho}\phi\phi D_{\sigma}\phi\psi \nn\\
   &+\frac{1}{2}\bar{\psi}D_{\mu}\phi\{F_{\alpha\nu},D_{\beta}\phi\}
   D_{\rho}\phi\phi D_{\sigma}\phi\psi \nn\\
   &+\frac{1}{2}\bar{\psi}D_{\mu}\phi D_{\nu}\phi\{F_{\alpha\rho},D_{\beta}
   \phi\}\phi D_{\sigma}\phi\psi \nn\\
   &+\frac{1}{2}\bar{\psi}D_{\mu}\phi D_{\nu}\phi 
D_{\rho}\phi\phi\{F_{\alpha\sigma},D_{\beta}\phi\}\psi\;\Bigg]\ .\label{mass2}
\end{align}
\begin{align}
   \widehat{S}^{(1)}_{m,3}=\frac{ic_{3}}{2}\Big(\frac{m}{l}
-\frac{2}{l^{2}}\Big)\;\theta^{\alpha\beta}\;\int 
d^{4}x\;
\varepsilon^{\mu\nu\rho\sigma}
\Bigg[&+\frac{i}{2}\bar{\psi}D_{\alpha}(D_{\mu}
\phi D_{\nu}\phi\phi D_{\rho}\phi D_{\sigma}\phi)D_{\beta}\psi \nn\\
   &-\frac{1}{4}\bar{\psi}F_{\alpha\beta}D_{\mu}\phi D_{\nu}\phi\phi 
   D_{\rho}\phi D_{\sigma}\phi\psi \nn\\
   &+\frac{i}{2}\bar{\psi}D_{\alpha}(D_{\mu}\phi D_{\nu}\phi\phi)
   D_{\beta}(D_{\rho}\phi D_{\sigma}\phi)\psi \nn\\
   &+\frac{i}{2}\bar{\psi}D_{\alpha}(D_{\mu}\phi 
   D_{\nu}\phi)D_{\beta}\phi D_{\rho}\phi D_{\sigma}\phi\psi\nn \\
  &+\frac{i}{2}\bar{\psi}(D_{\alpha}D_{\mu}\phi)(D_{\beta}D_{\nu}\phi)\phi D_{\rho}\phi D_{\sigma}\phi\psi \nn\\
   &+\frac{i}{2}\bar{\psi}D_{\mu}\phi D_{\nu}\phi\phi(D_{\alpha}D_{\rho}\phi)(D_{\beta}D_{\sigma}\phi)\psi \nn\\
   &+\frac{1}{2}\bar{\psi}\{F_{\alpha\mu},D_{\beta}\phi\}D_{\nu}\phi\phi 
   D_{\rho}\phi D_{\sigma}\phi\psi \nn\\
   &+\frac{1}{2}\bar{\psi}D_{\mu}\phi\{F_{\alpha\nu},D_{\beta}\phi\}\phi 
   D_{\rho}\phi D_{\sigma}\phi\psi\nn \\
   &+\frac{1}{2}\bar{\psi}D_{\mu}\phi D_{\nu}\phi\phi\{F_{\alpha\rho},D_{\beta}\phi\}D_{\sigma}\phi\psi \nn\\
   &+\frac{1}{2}\bar{\psi}D_{\mu}\phi D_{\nu}\phi\phi D_{\rho}\phi\{F_{\alpha\sigma},D_{\beta}\phi\}\psi\;\Bigg]\ .\label{mass3}
\end{align}

None of the three mass terms in (\ref{NC mass term}) is more preferable than 
others, and so we will treat all three of them on equal footing. 
According to (\ref{mass-afterSB-coefs}), we should assume that $c_1=-c_2=c_3$ if 
they are to contribute equally, at the classical level, to the Dirac mass term 
after the symmetry breaking. The coefficients must also satisfy the constraint 
(\ref{constraint}), and so we will set $c_1=-c_2=c_3=\tfrac{1}{72}$\ . 

\newpage
\noindent
After the symmetry breaking, the first order NC correction to the sum of the three mass terms becomes
\begin{eqnarray} 
   \widehat{S}^{(1)}_{m}=\theta^{\alpha\beta}\;\Bigg[&-&\frac{i}{4}\mass\int d^{4}x\;e\;(\nabla_{\alpha}e_{\mu}^{a})e^{\mu}_{a}\;\bar{\psi}\nabla_{\beta}\psi \nn \\
&+&\frac{1}{24}\mass\int d^{4}x\;e\;\eta_{ab}(\nabla_{\alpha}e_{\mu}^{a})(\nabla_{\beta}e_{\nu}^{b})\;
\bar{\psi}\sigma^{\mu\nu}\psi \nn \\
&-&\frac{1}{12}\mass\int d^{4}x\;e\;(\nabla_{\alpha}e_{\mu}^{a})(\nabla_{\beta}e_{\nu}^{b})(e^{\mu}_{a}e^{\nu}_{c}-e^{\mu}_{c}e^{\nu}_{a})\;\bar{\psi}\sigma^{c}_{\;\;b}\psi \nn \\
&-&\frac{1}{36}\massl\int d^{4}x\;e\;(\nabla_{\alpha}e_{\mu}^{a})e^{\mu}_{a}\;\bar{\psi}\gamma_{\beta}\psi \nn \\
&-&\frac{1}{96}\mass\int d^{4}x\;e\;R_{\alpha\beta}^{\;\;\;\;ab}\;\bar{\psi}\sigma_{ab}\psi \nn\\
&-&\frac{1}{12}\mass\theta^{\alpha\beta}\int d^{4}x\;e\;R_{\alpha\mu}^{\;\;\;\;ab}e^{\mu}_{a}e_{\beta}^{c}\;\bar{\psi}\sigma_{bc}\psi \nn \\
&-&\frac{1}{72}\massl\int d^{4}x\;e\;T_{\alpha\beta}^{\;\;\;\;a}\;\bar{\psi}\gamma_{a}\psi \nn\\
&-&\frac{7}{72}\massl\theta^{\alpha\beta}\int d^{4}x\;e\;T_{\alpha\mu}^{\;\;\;\;a}e^{\mu}_{a}\;\bar{\psi}
\gamma_{\beta}\psi \nn \\
&-&\frac{1}{8}\massll\int 
d^{4}x\;e\;\bar{\psi}\sigma_{\alpha\beta}\psi\;\;\;\Bigg]+h.c. \ . \label{total 
mass term after SB}
\end{eqnarray}

The full NC action at the first order in $\theta$ after the symmetry breaking is a sum of the kinetic term 
(\ref{kinetic-term-afterSB}) and the mass term (\ref{total mass term after SB}),
\be \widehat{S}^{(1)}=\widehat{S}^{(1)}_{kin}+\widehat{S}^{(1)}_{m}\ .\label{NC-action-afterSB}\ee 
The result (\ref{NC-action-afterSB}) is the desired first order noncommutative 
correction to the Dirac action in curved space-time. This action 
couples the Dirac bilinear quantities to the geometrical quantities like curvature, torsion, etc. It is manifestly $SO(1,3)$ gauge invariant and also, as we shall elaborate below, charge-conjugation invariant. The non-vanishing of the first order NC correction to the Dirac action in curved space-time is a significant result. Working in a different framework, it was argued in \cite{PLGR-fer} that this NC correction vanishes (along with all other odd-power corrections). However, in the framework presented here, in which gravity emerges as a consequence of gauge symmetry breaking and which is, at the same time, suitable for NC deformation, we have explicitly demonstrated by detailed calculation that the first order NC correction actually survives. This significant difference enables us to extract potentially observable NC effects already at the lowest perturbative order. We will see in the next section that the first order NC corrections to the Dirac action remain also in the limit of flat space-time, making it even easier to investigate modifications of e.g. Feynman propagator or the dispersion relation for electrons. 
Note that the first non-vanishing NC correction to the Einstein-Hilbert action 
is 
at the second order in $\theta$. This result is 
confirmed in many papers 
\cite{SWmapApproach,UsLetter,Us-16,MiAdSGrav,MDVR-14,PLGR-fer}.

\newpage

\subsection{C-conjugation}

Let us analyse the behaviour of the action (\ref{NC-action-afterSB}) under charge 
conjugation transformation. The charge conjugation operator is denoted by ${\mathcal C}$. The undeformed
Dirac field (now treated as an operator) and its adjoint transform in the following way:
\bea{\cal C}\psi(x){\cal C}^{-1}&=&-\bar{\psi}(x)C\ ,\nn\\
{\cal C}\bar\psi(x){\cal C}^{-1}&=&-{\psi}(x)C^{-1}\ ,
\eea
where $C$ is a matrix with the property $C\g_\m C^{-1}=-\g_\m^T$. Now we can easily find that, for example, 
\be {\cal C}\bar\psi\s_{\a\b}\psi{\cal C}^{-1}=-\bar\psi\s_{\a\b}\psi\ , \ee
and 
\be {\cal 
C}\Big(\bar\psi\g_a\nabla_\s\psi+(\nabla_\s\bar\psi)\g_a\psi\Big){\cal 
C}^{-1}=-\Big(\bar\psi\g_a\nabla_\s\psi+(\nabla_\s\bar\psi)\g_a\psi\Big) \ .\ee
For the rest of the terms in (\ref{NC-action-afterSB}) we obtain the 
similar result. 
This, together with the transformation law for the deformation parameter \cite{C-parity, Jabbari},
\be {\cal C}\theta^{\a\b}{\cal C}^{-1}=-\theta^{\a\b}\ , \label{-theta}\ee
leads to the conclusion that action (\ref{NC-action-afterSB}) is indeed invariant under charge conjugation. Such a behaviour for the deformation parameter $\theta^{\a\b}$ under charge conjugation transformation can be justified in several ways. For example, consider the transformation law for the NC Dirac spinor field $\widehat{\psi}$ and its adjoint. These fields can be expanded via Seiberg-Witten map as in (\ref{psi-expansion}) and (\ref{psibar-expansion}), and it must be ensured that they have the same sort of behaviour under C-conjugation as their undeformed counterparts, i.e. we demand that
\bea{\cal C}\widehat{\psi}(x){\cal C}^{-1}&=&-\widehat{\bar{\psi}}(x)C\ ,\nn\\
{\cal C}\widehat{\bar\psi}(x){\cal C}^{-1}&=&-\widehat{\psi}(x)C^{-1}\ . \label{NC C-conjug}
\eea
Using the $C\o_{\a}C^{-1}=-\o_{\a}^{T}$ property of the gauge potential, we can readily deduce
\bea
{\cal C}(\o_{\a}\partial_{\b}\psi){\cal C}^{-1}&=&\partial_{\b}\bar{\psi}(\o_{\a}C) \ , \\
{\cal C}(\o_{\a}D_{\b}\psi){\cal C}^{-1}&=&D_{\b}\bar{\psi}(\o_{\a}C)\ ,
\eea 
which are the transformation properties of the terms appearing in (\ref{psi-expansion}).
If we want (\ref{NC C-conjug}) to hold, $\theta^{\a\b}$ must change its sign under C-conjugation. Aside from this formal argument based on symmetry considerations, there is a heuristic argument for assuming the transformation law (\ref{-theta}) given in 
\cite{Jabbari}, stemming from string theory. It is explained there that an electric dipol momentum of an open string is proportional to $\theta^{\a\b}$. This motivates the conclusion that $\theta^{\a\b}$ goes to $-\theta^{\a\b}$ under charge conjugation transformation. However, there is no contradiction with the original definition of $\theta^{\a\b}$ given in (\ref{can-kom-rel}), where it is understood as a measure of incompatibility of space-time coordinates, because the Moyal-Weyl $\star$-product,  defined in (\ref{moyal}), also changes under the charge conjugation, namely, $\star_{\theta}\rightarrow \star_{-\theta}$. In other words, charge conjugation transformation does not change the noncommutative algebra of space-time coordinates.   

\initiate
\section{NC Dirac equation in flat space-time}

In this last section, we study the special case of flat space-time in 
order to investigate the influence of noncommutativity (which survives in this limit) on the energy-momentum relation for electrons. In the flat space-time limit, the action (\ref{NC-action-afterSB}) becomes 
\begin{align}
   \widehat{S}^{(1)}=\theta^{\alpha\beta}\int d^{4}x\;\Bigg[-\frac{1}{2l}\bar{\psi}
\sigma_{\alpha}^{\;\;\sigma}\partial_{\beta}\partial_{\sigma}\psi+\frac{7i}{24l^ {2}}\varepsilon_{\alpha\beta}^{\;\;\;\;\rho\sigma}\bar{\psi}
\gamma_{\rho}\gamma_ 
{5}\partial_{\sigma}\psi-M\bar\psi\sigma_{\alpha\beta}\psi\Bigg]\ ,\label{Dirac-action-flat-first-corrcetion}
\end{align}
where we introduced the notation $M:=\frac{m}{4l^{2}}+\frac{1}{6l^{3}}$. 
Therefore, the effect of noncommutativity, in the form of new couplings in the action, is relevant even in flat 
space-time.\\[0.2cm]
The total NC action in flat space-time to the first order is
\begin{align}
   \widehat{S}=\widehat{S}^{(0)}+\widehat{S}^{(1)}=&\int d^{4}x\;\bar{\psi}(i\gamma^{\m}\partial_{\m}-m)\psi
 +\theta^{\alpha\beta}\int d^{4}x\;\Bigg[-\frac{1}{2l}\bar{\psi}
\sigma_{\alpha}^{\;\;\sigma}\partial_{\beta}\partial_{\sigma}\psi \nn \\ 
&\;\;\;\;\;\;\;\;\;\;\;\;\;\;\;\;\;\;\;+\frac{7i}{24l^ {2}}\varepsilon_{\alpha\beta}^{\;\;\;\;\rho\sigma}\bar{\psi}
\gamma_{\rho}\gamma_ 
{5}\partial_{\sigma}\psi-M\bar\psi\sigma_{\alpha\beta}\psi\Bigg]\ .\label{Dirac-action-flat}
\end{align}
The existence of the first order NC correction to the Dirac action is a 
nontrivial consequence of the model presented here. We can easily derive the 
Feynman propagator for the Dirac field from the action (\ref{Dirac-action-flat}). 
The result (in momentum space) is given by
\bea 
iS_{F}(p)&=&\int d^4x\; \langle{\Omega}|T\psi(x)
\bar\psi(0)|{\Omega}\rangle e^{ipx}\nn\\
&=&\frac{i}{\slash 
p-m+i\epsilon}+\frac{i}{\slash 
p-m+i\epsilon}(i\theta^{\a\b}D_{\a\b})\frac{i}{\slash p-m+i\epsilon}+\dots\ ,
\eea
where 
\be D_{\a\b}:=\frac{1}{2l}
\sigma_{\alpha}^{\;\;\sigma}p_{\beta}p_{\sigma}  
+\frac{7}{24l^ {2}}\varepsilon_{\alpha\beta}^{\;\;\;\;\rho\sigma}
\gamma_{\rho}\gamma_ 
{5}p_{\sigma}-M\sigma_{\alpha\beta}\ .
\ee
The Feynman propagator is modified due to the space-time noncommutativity. Thus, we see 
that an electron effectively interacts with the NC background itself. In this respect, we may say that NC background acts like a background electromagnetic field. \\[0.2cm] 
\noindent
By varying action (\ref{Dirac-action-flat}) with respect to 
$\bar{\psi}$ we derive the modified Dirac equation in Minkowski space-time:
\be
  \Bigg[i\slashed{\partial}-m-\frac{1}{2l}\theta^{\alpha\beta}
  \sigma_{\alpha}^{\;\;\sigma}\partial_{\beta}\partial_{\sigma}+\frac{7i}{24l^{2}}\theta^{\alpha\beta}
\varepsilon_{\alpha\beta}^{\;\;\;\;\rho\sigma}\gamma_{\rho}
\gamma_{5}\partial_{\sigma}
-\theta^{\alpha\beta}M\sigma_{\alpha\beta}\Bigg]\psi=0\ .\label{Deq-nc}
\ee
 \\[0.2cm]
To simplify further analysis, we will assume that only two spatial dimensions are mutually incompatible, e.g. $[x^{1}, x^{2}]=i\theta^{12}$. Thus, we have $\theta^{12}=-\theta^{21}=:\theta\neq0$ and all other components of $\theta^{\m\n}$ equal to zero. The equation (\ref{Deq-nc}) reduces to
 \be
  \Bigg[i\slashed{\partial}-m-\frac{\theta}{2l}(\sigma_{1}^{\;\;\sigma}\partial_{2}\partial_{\sigma}-\sigma_{2}^{\;\;\sigma}\partial_{1}\partial_{\sigma})+\frac{7i\theta}{12l^{2}}
(\gamma_{0}\gamma_{5}\partial_{3}-\gamma_{3}\gamma_{5}\partial_{0})-2\theta
M\sigma_{12}\Bigg]\psi=0\ ,\label{NCDirac-flat-eq}
\ee
where we assumed the convention in which $\varepsilon^{0123}=1$.\\[0.2cm] 
Let us now find the dispersion relation, i.e. energy-momentum relation, for the Dirac fermions. We will assume the plane wave ansatz $\psi(x)=u({\bf p})e^{-i
p\cdot x}$ where $u(\textbf{p})$ stands for a yet undetermined spinor amplitude
\be
u(\textbf{p})=\begin{pmatrix}
  a   \\[6pt]
  b   \\[6pt]
  c   \\[6pt]
  d     
     \end{pmatrix}\ .
\ee
With this choice, equation (\ref{NCDirac-flat-eq}) can be represented in the momentum space as
\be
\Bigg(\begin{pmatrix}E-m&-{\boldsymbol{\sigma}}\cdot{\bf p}\\
{\boldsymbol{\s}}\cdot{\bf p}&-E-m
\end{pmatrix}
+\theta {\cal M}\Bigg)u({\textbf{p}})=0\ ,\label{NCDirac-free-impulsp}\ee 
where the matrix ${\cal M}$ is given by 
\be{\cal M}=
\begin{pmatrix}
  A&   \frac{1}{2l}
  p_{z}p_{-} &  -\frac{7}{12l^2}p_{z} & \frac{1}{2l}Ep_{-}  \\[6pt]
\frac{1}{2l}p_{z}p_{+} & -A
     &  -\frac{1}{2l}Ep_{+} & -\frac{7}{12l^2}p_{z}\\[6pt]
     \frac{7}{12l^2}p_{z} &  \frac{1}{2l}Ep_{-} & B & \frac{1}{2l}p_{z}p_{-}  
  \\[6pt]
     -\frac{1}{2l}Ep_{+} & \frac{7}{12l^2}p_{z} &   \frac{1}{2l}p_{z}p_{+}  &
   -B    
     \end{pmatrix}\ .
\ee
Quantities $E$ and ${\bf p}$ denote energy and momentum of a particle, respectively, and
the matrix elements $A$ and $B$ are given by
\bea
A&:=&-\frac{1}{2l}( p^{2}_{x}+p_{y}^2)+\frac{7E}{12l^2}-2M\ , \nn \\
B&:=&-\frac{1}{2l}( p^{2}_{x}+p_{y}^2)-\frac{7E}{12l^2}-2M\ ,\eea
with $p_{\pm}=p_{x}\pm ip_{y}$. 
We use the Dirac representation of $\gamma-$matrices. \\[0.2cm] 
Nontrivial solutions of the homogeneous matrix equation (\ref{NCDirac-free-impulsp}) which, when written explicitly, states that     
\be
\begin{pmatrix}
  E-m+\theta A  & \frac{\theta}{2l}
  p_{z}p_{-} &  -p_{z}-\frac{7\theta}{12l^2}p_{z} & -p_{-}+\frac{\theta}{2l}Ep_{-}  \\[6pt]
\frac{\theta}{2l}p_{z}p_{+} & E-m-\theta A
     &  -p_{+}-\frac{\theta}{2l}Ep_{+} & p_{z}-\frac{7\theta}{12l^2}p_{z}\\[6pt]
     p_{z}+\frac{7\theta}{12l^2}p_{z} &  p_{-}+\frac{\theta}{2l}Ep_{-} &-E-m+\theta B & \frac{\theta}{2l}p_{z}p_{-}  
  \\[6pt]
     p_{+}-\frac{\theta}{2l}Ep_{+} &-p_{z}+\frac{7\theta}{12l^2}p_{z} & \frac{\theta}{2l}p_{z}p_{+}  &
  -E-m-\theta B    
     \end{pmatrix}\begin{pmatrix}
  a   \\[6pt]
  b   \\[6pt]
  c   \\[6pt]
  d     
     \end{pmatrix}=0 \ ,\label{det=0} \\
\ee
exist if and only if the 
determinant of the matrix $\slash p -m+\theta\cal M$ (which is the matrix appearing in (\ref{det=0})) equals zero. This condition will give us the dispersion relation. Since we are working with the NC action up to the first order in $\theta$, the result for the determinant is reliable only up to this order. The determinant depends on the energy which is also represented as a perturbative expansion in powers of $\theta$, 
\be
E=\sum_{n=0}^{+\infty}E^{(n)}\ ,\;\;\text{where}\;\;E^{(n)}\sim\frac{\theta^{n}}{(\text{lenght})^{2n+1}}\ .\label{energy-expansion}
\ee
If the determinant is equal to zero, it is equal to zero order by order in $\theta$, and we can derive the momentum dependence of $E^{(1)}$ term in the energy expansion, which is enough to see how noncommutativity influences the dispersion relation for the Dirac fermions. To get higher order energy terms we need higher order perturbative corrections to the Dirac action. \\[0.2cm]
First we will consider an electron moving along the $z$-direction, i.e. in the direction orthogonal to the noncommutative $x,y$-plane. In this case, matrix equation (\ref{det=0}) reduces to 
\be
\begin{pmatrix}
  E-m+\theta A(0)  & 0 &  -p_{z}-\frac{7\theta}{12l^2}p_{z} & 0  \\[6pt]
0 & E-m-\theta A(0)
     & 0 & p_{z}-\frac{7\theta}{12l^2}p_{z}\\[6pt]
     p_{z}+\frac{7\theta}{12l^2}p_{z} & 0 &-E-m+\theta B(0) & 0  
  \\[6pt]
     0 &-p_{z}+\frac{7\theta}{12l^2}p_{z} & 0  &
  -E-m-\theta B(0)    
     \end{pmatrix}\begin{pmatrix}
  a   \\[6pt]
  b   \\[6pt]
  c   \\[6pt]
  d     
     \end{pmatrix}=0 \ , \label{det=0-pz}
\ee
where $A(0)=A(p_{x}=p_{y}=0)$ and likewise $B(0)=B(p_{x}=p_{y}=0)$. \\[0.2cm]
We can equivalently write the following system of equations: 
\bea
\left[E-m+\left(\frac{7E}{12l^2}-2M\right)\theta\right]a-\left[p_{z}+\frac{7p_{z}}{12l^2}\theta\right]c &=& 0 \ , \nn \\
\left[p_{z}+\frac{7p_{z}}{12l^2}\theta\right]a-\left[E+m+\left(\frac{7E}{12l^2} +2M\right)\theta\right]c &=& 0 \ , \nn \\
\left[E-m-\left(\frac{7E}{12l^2} -2M\right)\theta\right]b+\left[p_{z}-\frac{7p_{z}}{12l^2}\theta\right]d &=& 0 \ , \nn \\
\left[-p_{z}+\frac{7p_{z}}{12l^2}\theta\right]b+\left[-E-m+\left(\frac{7E}{12l^2} +2M\right)\theta\right]d &=& 0 \ . \label{system-pz}
\eea
The first pair of equations is decoupled from the remaining two and we can treat those two pairs separately. 
 Non trivial solution for spinor components $a$, $b$, $c$, and $d$ exist if at least one of the following two conditions is satisfied:
\be
\left[E-m\pm\left(\frac{7E}{12l^2}-2M\right)\theta\right]\left[E+m\pm\left(\frac{7E}{12l^2}+2M\right)\theta\right]=\left[p_{z}\pm\frac{7p_{z}}{12l^2}\theta\right]^{2} \ .
\label{system2} \ee 
We finally obtain four different solutions for the energy (taken to the first order in $\theta$, i.e. $E=E^{(0)}+E^{(1)}$):
\bea
E_{1,2}&=&E_{\textbf{p}}\mp\left[\frac{m^{2}}{12l^{2}}-\frac{m}{3l^{3}}\right]\frac{\theta}{E_{\textbf{p}}}+\mathcal{O}(\theta^{2}) \ , \nn \\
E_{3,4}&=&-E_{\textbf{p}}\pm\left[\frac{m^{2}}{12l^{2}}-\frac{m}{3l^{3}}\right]\frac{\theta}{E_{\textbf{p}}}+\mathcal{O}(\theta^{2}) \ , \label{energy-solutions-pz}
\eea
with $E_{\textbf{p}}=\sqrt{m^{2}+p_{z}^{2}}$. This is a reminiscent of the well known Zeeman effect. The deformation parameter $\theta$ plays the role of a constant background magnetic field that causes the splitting of atomic energy levels. In the rest frame the energies reduce to:
\bea
E_{1,2}(0)&=&m\mp\left[\frac{m}{12l^{2}}-\frac{1}{3l^{3}}\right]\theta+\mathcal{O}(\theta^{2}) \ , \nn \\
E_{3,4}(0)&=&-m\pm\left[\frac{m}{12l^{2}}-\frac{1}{3l^{3}}\right]\theta+\mathcal{O}(\theta^{2}) \ . \label{energy-solutions}
\eea
From (\ref{energy-solutions}) we see that the electron's mass gets renormalised due to the noncommutativity of the background space-time. The correction is linear in the deformation parameter which is a significant and curious result. To our knowledge, this aspect of noncommutativity has not been discovered in the previous studies of the subject. \\[0.2cm]
By solving the 
system (\ref{system-pz}) for each of the four energy functions in (\ref{energy-solutions-pz}), we get the following four linearly independent solutions of the NC Dirac equation (up to a normalization factor):
\bea
& \psi_{1}\sim\begin{pmatrix}
  1   \\[6pt]
  0   \\[6pt]
  \frac{p_{z}}{E_{\textbf{p}}+m}\left[1+\left(\frac{m}{12l^{2}}-\frac{1}{3l^{3}}\right)\frac{\theta}{E_{\textbf{p}}}\right] &  \\[6pt]
  0     
     \end{pmatrix}e^{-iE_{1}t+ip_{z}z}\ , \nn \\[0.4cm]   
    & \psi_{2}\sim\begin{pmatrix}
  0   \\[6pt]
  1   \\[6pt]
  0   \\[6pt]
  \frac{p_{z}}{E_{\textbf{p}}+m}\left[1-\left(\frac{m}{12l^{2}}-\frac{1}{3l^{3}}\right)\frac{\theta}{E_{\textbf{p}}}\right]   &   
     \end{pmatrix}e^{-iE_{2}t-ip_{z}z}\ , \nn \\[0.4cm]
& \psi_{3}\sim\begin{pmatrix}
 \frac{p_{z}}{E_{\textbf{p}}+m}\left[1+\left(\frac{m}{12l^{2}}-\frac{1}{3l^{3}}\right)\frac{\theta}{E_{\textbf{p}}}\right]  &   \\[6pt]
  0   \\[6pt]
  1   \\[6pt]
  0     
     \end{pmatrix}e^{-iE_{3}t-ip_{z}z}\ , \nn \\[0.4cm]
   &  \psi_{4}\sim \begin{pmatrix}
  0   \\[6pt]
  \frac{p_{z}}{E_{\textbf{p}}+m}\left[1-\left(\frac{m}{12l^{2}}-\frac{1}{3l^{3}}\right)\frac{\theta}{E_{\textbf{p}}}\right]   &   \\[6pt]
  0   \\[6pt]
  1     
     \end{pmatrix}e^{-iE_{4}t+ip_{z}z}\ .     
\eea 
Spinors $\psi_{1}$ and $\psi_{2}$ ($\psi_{3}$, and $\psi_{4}$) correspond to positive (negative) frequency solutions of the NC Dirac equation to the first order. Let us note that in commutative (classical) case the opposite helicity $(\pm\frac{1}{2})$ solutions have the same energy. However, in noncommutative case, as we can see, the solutions with opposite helicity have different energies. The noncommutativity of space, here taken to be confined in $x,y$-plane, causes the classical energy levels $\pm E_{\textbf{p}}$ to split. The energy gap between the new-formed levels is the same for $E_{\textbf{p}}$ and $-E_{\textbf{p}}$ and it is equal to 
\be
2\left[\frac{m^{2}}{12l^{2}}-\frac{m}{3l^{3}}\right]\frac{\theta}{E_{\textbf{p}}}\ .
\ee        
From the dispersion relations (\ref{energy-solutions-pz}) we can easily find the (group) velocity of an electron. This velocity is defined by
\be
\textbf{v}\equiv\frac{\partial E}{\partial \textbf{p}}\ .
\ee
For positive (negative) helicity solution $\psi_{1}$ $(\psi_{2})$ we get 
\be
\textbf{v}_{1,2}=\frac{\textbf{p}}{E_{\textbf{p}}}\left[1\pm\left(\frac{m^{2}}{12l^{2}}-\frac{m}{3l^{3}}\right)\frac{\theta}{E_{\textbf{p}}^{2}}+\mathcal{O}(\theta^{2})\right]\ .
\ee	
These velocities can be rewritten in the following way:\bea
\textbf{v}_{1,2}=\frac{\textbf{p}}{E_{1,2}}+\mathcal{O}(\theta^{2})\ .
\eea 
Thus, we conclude that velocity of an electron moving in $z$-direction depends on its helicity. This is analogues to the birefringence effect, i.e. the optical property of a material having a refractive index that depends on the polarization and propagation direction of light. \\[0.2cm]
The Dirac spinor $\psi_{1}$ can be represented as
\be
\psi_{1}\sim\begin{pmatrix}
  1   \\[6pt]
  0   \\[6pt]
  \frac{p_{z}}{E_{1}+E_{1}(0)} &  \\[6pt]
  0     
     \end{pmatrix}e^{-iE_{1}t+ip_{z}z}\ .
\ee
and the corresponding Dirac spinor in the rest frame is 
\be
\psi_{1}(0)\sim  \begin{pmatrix}
  1   \\[6pt]
  0   \\[6pt]
  0   \\[6pt]
  0     
     \end{pmatrix}e^{-iE_{1}(0)t} \ ,
   \ee
where $E_{1}(0)$ is the value of energy $E_{1}$ for $p_{z}=0$,
\be
E_{1}(0)=m-\left[\frac{m}{12l^{2}}-\frac{1}{3l^{3}}\right]\theta\ . 
\ee
The boost along $z$-direction in spinor representation is given by
\be
S(\varphi)=\cosh\left(\frac{\varphi}{2}\right)I-\sinh\left(\frac{\varphi}{2}\right)\begin{pmatrix}
  0 & \sigma_{3} \\[6pt]
  \sigma_{3} & 0  
     \end{pmatrix}\ , \label{boost-z}
\ee    
where $v=\tanh(\varphi)$. If we take $v=-v_{1}=-\tfrac{p_{z}}{E_{1}}$ we can construct the boost matrix that transforms the rest frame solution $\psi_{1}(p_{z}=0)$ into the solution $\psi_{1}(p_{z})$. It is given by
\be
S(-p_{z})=\sqrt{\frac{E_{1}(p_{z})+E_{1}(0)}{2E_{1}(0)}}I+\sqrt{\frac{E_{1}(p_{z})-E_{1}(0)}{2E_{1}(0)}}\begin{pmatrix}
  0 & \sigma_{3} \\[6pt]
  \sigma_{3} & 0  
     \end{pmatrix}\ ,
\ee        
and we have
\be
S(-p_{z})\psi_{1}(0)=\psi_{1}(p_{z})\ .
\ee
This result shows that constant noncommutativity in $x,y$-plane is compatible with the Lorentz boosts along $z$-direction. The similar statement holds for the other solutions. 
\newpage
\noindent
For an electron moving in noncommutative $x,y$-plane, i.e. an electron whose momentum is $\textbf{p}=(p_{x},p_{y},0)$, by using the same procedure, we get the deformed energy levels:
\bea
E_{1,4}&=&\pm E_{\textbf{p}}-\left[\frac{m}{12l^{2}}-\frac{1}{3l^{3}}\right]\theta \ , \nn \\
E_{2,3}&=&\pm E_{\textbf{p}}+\left[\frac{m}{12l^{2}}-\frac{1}{3l^{3}}\right]\theta \ , \label{energy-solutions-px,py}
\eea
with $E_{\textbf{p}}=\sqrt{m^{2}+p_{x}^{2}+p_{y}^{2}}$. It is interesting to note that, in this case, NC corrections do not depend on the momentum, as opposed to the NC corrections of the energy levels of an electron moving along $z$-direction, i.e. in the direction in which it does not feel the noncommutativity. Again, these energy levels exactly reduces to (\ref{energy-solutions}) when $\textbf{p}=0$. The four independent Dirac spinors are:\\[0.2cm]
\bea
& \psi_{1}\sim\begin{pmatrix}
  1  \\[6pt]
  0   \\[6pt]
  0   \\[6pt]
  \frac{p_{+}}{E_{\textbf{p}}+m}\left[1+\left(\frac{7}{12l^{2}}-\frac{m}{12l}\right)\theta\right] &   \\[6pt]  
     \end{pmatrix}e^{-iE_{1}t+ip_{x}x+ip_{y}y}\ , \nn \\[0.6cm]
 & \psi_{2}\sim\begin{pmatrix}
  0   \\[6pt]
  1   \\[6pt]
  \frac{p_{-}}{E_{\textbf{p}}+m}\left[1-\left(\frac{7}{12l^{2}}-\frac{m}{12l}\right)\theta\right]     &   \\[6pt]
   0    
     \end{pmatrix}e^{-iE_{2}t+ip_{x}x+ip_{y}y}\ , \nn \\[0.6cm]
& \psi_{3}\sim\begin{pmatrix}
  0 \\[6pt] 
 \frac{p_{+}}{E_{\textbf{p}}+m}\left[1+\left(\frac{7}{12l^{2}}-\frac{m}{12l}\right)\theta\right]  &   \\[6pt]
  1   \\[6pt]
  0     
     \end{pmatrix}e^{-iE_{3}t-ip_{x}x-ip_{y}y}\ , \nn \\[0.6cm]
 & \psi_{4}\sim \begin{pmatrix}
 \frac{p_{-}}{E_{\textbf{p}}+m}\left[1-\left(\frac{7}{12l^{2}}-\frac{m}{12l}\right)\theta\right]   &   \\[6pt]
  0   \\[6pt] 
  0   \\[6pt]
  1     
     \end{pmatrix}e^{-iE_{4}t-ip_{x}x-ip_{y}y}\ .     
\eea 
It turns out that these solutions cannot be obtained by boosting the corresponding rest frame solutions. This was to be expected since, as we have already mentioned, by choosing the canonical noncommutativity we have effectively fixed the coordinate system. In other words, we work in a preferred coordinate system in which only boosts along $z$-axis and rotations around $z$-axis are preserved.            

\newpage
\section{Conclusion}

\hspace{0.4cm} We have studied the coupling of the Dirac spinor field and gravity on 
the noncommutative Moyal-Weyl space-time. 
First, we introduced a commutative model based on $SO(2,3)$ gauge 
symmetry. Breaking this symmetry down to the Lorentz $SO(1,3)$ gauge symmetry 
reduces the action to the Dirac action in curved space-time. Then we deformed 
our classical model, thus obtaining its noncommutative generalisation. The NC 
action exhibits deformed $SO(2,3)_\star$ gauge symmetry. 
Assuming that deformation parameter $\theta^{\a\b}$ is small enough, we expanded the NC action perturbatively up to the 
first order in $\theta^{\a\b}$.
This is done by utilising the Seiberg-Witten map and the enveloping algebra 
approach. After the symmetry breaking, the obtained first order NC correction does 
not vanish. It has local Lorentz symmetry and it is invariant under the charge 
conjugation. We have noted that by introducing the canonical commutation 
relations (\ref{can-kom-rel}) we effectively fixed the coordinate system. In 
this way the diffeomorphism symmetry is manifestly broken. The coordinates in 
which we work are not arbitrary ones, but rather they are Fermi normal 
coordinates, as it was shown in \cite{UsLetter, Us-16}. In the last section we considered the limit of flat space-time. There is a modification of the Dirac equation 
and the Feynman propagator due 
to noncommutativity that appears already at the first order in 
$\theta^{\a\b}$. This result is different from the one that is obtained by directly introducing noncommutativity into the free Dirac action (minimal 
substitution) giving
\be S=\int 
d^{4}x\;\widehat{\bar{\psi}}\star(i\gamma^{\m}\partial_{\m}-m)\widehat{\psi}\ 
.\label{Dirac-action-minimal}\ee  
Since \be \int d^4 x\; \widehat{f}\star\widehat{g}=\int d^4x\;fg,\ee the first 
order NC correction of the free Dirac action (\ref{Dirac-action-minimal}) 
vanishes. This feature of our model leads to some significant consequences. The dispersion relation for electrons is modified due to noncommutativity already at the first order in $\theta^{\a\b}$. The undeformed energy levels of the classical theory are split in background NC space-time. We also found the explicit solutions of the NC Dirac equation in flat space-time and demonstrated that by introducing constant noncommutativity in flat space-time we are effectively working in the preferred coordinate systems. \\[0.2cm]
Let us mention the appearance of the term 
$\theta^{\a\b}\bar{\psi}\sigma_{\a\b}\psi$ in the NC Lagrangian density. It 
resembles the magnetic moment term in Electrodynamics with 
electromagnetic field strength tensor replaced by $\theta^{\a\b}$. If we 
interpret the parameter of noncommutativity as a constant 
''electric/magnetic'' background field the analogy becomes obvious. This is in accord with the behaviour of the deformation parameter under charge-conjugation and the upper mentioned Zeeman-like splitting of the classical energy levels. \\[0.2cm]
In future work we plan to include electromagnetic field in our NC $SO(2,3)_\star$ model. This will lead us to a theory of NC electrodynamics 
which can be compared to the one established by the standard approach based on 
the minimal substitution. Minimal NC electrodynamics is not a renormalisabile 
theory because of the fermionic loop contributions \cite{minimalNC-Eld}. It 
would be interesting to analyse the renormalisability of the presented model and 
to extend this approach to scalar and non-Abelian gauge fields in order to establish a complete theory concerning the behaviour of matter in noncommutative space-time.      

\vskip1cm \noindent 
{\bf Acknowledgement}
\hskip0.3cm
We would like to thank  Maja Buri\' 
c, Marija Dimitrijevi\'c-\'Ciri\'c and Aleksandra Dimi\'c 
for
fruitful discussion and useful comments. The work is
supported by project
ON171031 of the Serbian Ministry of Education and Science and partially
supported  by  the  Action  MP1405  QSPACE from  the  Europe
an  Cooperation  in  Science
and  Technology  (COST). V.R. would like to thank ESI (Vienna) for hospitality 
during his stay.    

\newpage
\appendix

\renewcommand{\theequation}{\Alph{section}.\arabic{equation}}
\initiate

\section{The Kinetic term}
The kinetic action term (\ref{NCaction-beforeSB}) contains eight terms before symmetry braking. 
We label them with an index according to the order of appearance in 
(\ref{NCaction-beforeSB}). After the symmetry breaking they are given by:  
\bea
  \widehat{S}^{(1)}_{1}=&+&\frac{1}{16}\theta^{\alpha\beta}\int d^{4}x\;e\;R_{\alpha\beta}^{\;\;\;\;ab}e^{\sigma}_{b}\;\bar{\psi}\gamma_{a}\nabla_{\sigma}\psi
  -\frac{i}{32}\theta^{\alpha\beta}
  \int d^{4}x\;e\;R_{\alpha\beta}^{\;\;\;\;ab}\varepsilon_{abc}^{\;\;\;\;\;\;d}e^{\sigma} _{d}\;\bar{\psi}\gamma^{c}\gamma^{5}\nabla_{\sigma}\psi\nn\\
  &+&\frac{i}{16l}\theta^{\alpha\beta}
  \int d^{4}x\;e\;T_{\alpha\beta}^{\;\;\;\;a}e^{\sigma}_{a}\;\bar{\psi}\nabla_{\sigma}\psi
  +\frac{1}{16l}\theta^{\alpha\beta}
  \int d^{4}x\;e\; T_{\alpha\beta}^{\;\;\;\;a}e^{\mu}_{a}\;\bar{\psi}\sigma_{\mu}^{\;\;\sigma}\nabla_{\sigma}\psi\nn\\
&+&\frac{i}{16l^{2}}\theta^{\alpha\beta}
\int d^{4}x\;e\;\varepsilon_{abc}^{\;\;\;\;\;\;d}e_{\alpha}^{a}e_{\beta}^{b}e^{\sigma}_{d}\;\bar{\psi}\gamma^{c}
\gamma^{5}\nabla_{\sigma}\psi
-\frac{1}{8l^{2}}\theta^{\alpha\beta}\int d^{4}x\;e\;\bar{\psi}\gamma_{\alpha}\nabla_{\beta}\psi \nn\\
&+&\frac{1}{16l}\theta^{\alpha\beta}\int d^{4}x\;e\;
R_{\alpha\beta}^{\;\;\;\;ab}\;\bar{\psi}\sigma_{ab}\nn\psi\\
&-&\frac{1}{8l^{2}}\theta^{\alpha\beta}\int d^{4}x\;e\;
T_{\alpha\beta}^{\;\;\;\;a}\;\bar{\psi}\gamma_{a}\psi\nn\\
&-&\frac{1}{8l^{3}}\theta^{\alpha\beta}\int d^{4}x\;e\;\bar{\psi}\sigma_{\alpha\beta}\psi
\eea

\bea
  \widehat{S}_{2}^{(1)}=&-&\frac{1}{4}\theta^{\alpha\beta}\int d^{4}x\;e\;(\nabla_{\alpha}e_{\mu}^{a})(e^{\mu}_{a}e^{\sigma}_{b}-e^{\mu}_{b}e^{\sigma}_{a})\;\bar{\psi}\gamma^{b}\nabla_{\beta}\nabla_{\sigma}\psi
  -\frac{1}{4l}\theta^{\alpha\beta}\int d^{4}x\;e\;\bar{\psi}\sigma_{\alpha}^{\;\;\sigma}\nabla_{\beta}
  \nabla_{\sigma}\psi\nn\\
  &-&\frac{i}{2l}\theta^{\alpha\beta}\int d^{4}x\;e\;(\nabla_{\alpha}e_{\mu}^{a})e^{\mu}_{a}\;\bar{\psi}\nabla_{\beta}\psi
  -\frac{1}{8l}\theta^{\alpha\beta}\int d^{4}x\;e\;(\nabla_{\alpha}e_{\mu}^{a})e^{\mu}_{b}\;\bar{\psi}\sigma_{a}^{\;\;b}
  \nabla_{\beta}\psi \nn \\
  &+&\frac{1}{2l^{2}}\theta^{\alpha\beta}\int d^{4}x\;e\;\bar{\psi}\gamma_{\alpha}\nabla_{\beta}\psi
  +\frac{i}{16l}\theta^{\alpha\beta}\int d^{4}x\;e\;
  T_{\alpha\beta}^{\;\;\;\;a}e^{\sigma}_{a}\;\bar{\psi}\nabla_{\sigma}\psi \nn \\
  &-&\frac{1}{8l}\theta^{\alpha\beta}\int d^{4}x\;e\;(\nabla_{\alpha}e_{\mu}^{a})(e^{\mu}_{a}e^{\sigma}_{b}-e^{\mu}_{b}e^{\sigma}_{a})e_{\beta}^{c}\;\bar{\psi}\sigma^{b}_{\;\;c}
  \nabla_{\sigma}\psi\nn\\
  &+&\frac{i}{8l^{2}}\theta^{\alpha\beta}\int d^{4}x\;e\;\varepsilon_{ab}^{\;\;\;\;cd}e_{\alpha}^{a}e_{\beta}^{b}e^{\sigma}_{d}\;\bar{\psi}\gamma_{c}\gamma_{5}
  \nabla_{\sigma}\psi \nn\\
  &-&\frac{1}{8l}\theta^{\alpha\beta}\int d^{4}x\;e\;(\nabla_{\alpha}e_{\mu}^{a})(\nabla_{\beta}e_{\nu}^{b})
  (e^{\mu}_{a}e^{\nu}_{c}-e^{\mu}_{c}e^{\nu}_{a})\;\bar{\psi}\sigma^{c}_{\;\;b}\psi \nn \\
    &+&\frac{i}{16l^{2}}\theta^{\alpha\beta}\int d^{4}x\;e\;(\nabla_{\alpha}e_{\mu}^{a})e^{\mu}_{b}e_{\beta}^{c}
    \varepsilon_{a\;\;cd}^{\;\;b}\;\bar{\psi}\gamma^{d}\gamma_{5}\psi \nn\\ 
  &-&\frac{3}{16l^{2}}\theta^{\alpha\beta}\int d^{4}x\;e\;(\nabla_{\alpha}e_{\mu}^{a})e^{\mu}_{a}\;\bar{\psi}\gamma_{\beta}\psi \nn\\
  &+&\frac{1}{16l^{2}}\theta^{\alpha\beta}\int d^{4}x\;e\;(\nabla_{\alpha}e_{\mu}^{a})e_{\beta a}\;\bar{\psi}\gamma^{\mu}\psi\nn \\
  &-&\frac{1}{32l^{2}}\theta^{\alpha\beta}\int d^{4}x\;e\;T_{\alpha\beta}^{\;\;\;\;a}\;\bar{\psi}\gamma_{a}\psi\nn\\
  &-&\frac{1}{16l^{3}}\theta^{\alpha\beta}\int d^{4}x\;e\;\bar{\psi}\sigma_{\alpha\beta}\psi
\eea

\bea
  \widehat{S}_{3}^{(1)}=&+&\frac{i}{12}\theta^{\alpha\beta}\int d^{4}x\;e\;(\nabla_{\alpha}e_{\mu}^{a})(\nabla_{\beta}e_{\nu}^{b})\varepsilon_{b}^{\;\;cds}e_{c}^{\mu}e_{d}^{\nu}e_{s}^{\sigma}\;\bar{\psi}\gamma_{a}\gamma_{5}\nabla_{\sigma}\psi \nn\\
&-&\frac{i}{12}\theta^{\alpha\beta}\int d^{4}x\;e\;\eta_{ab}(\nabla_{\alpha}e_{\mu}^{a})(\nabla_{\beta}e_{\nu}^{b})\varepsilon^{cdrs}e_{c}^{\mu}e_{d}^{\nu}e_{s}^{\sigma}\;\bar{\psi}\gamma_{r}
\gamma_{5}\nabla_{\sigma}\psi \nn\\
&-&\frac{1}{12l}\theta^{\alpha\beta}\int d^{4}x\;e\;(\nabla_{\beta}e_{\nu}^{b})\varepsilon_{b}^{\;\;cds}e_{\alpha c}e_{d}^{\nu}e_{s}^{\sigma}\;\bar{\psi}\gamma_{5}\nabla_{\sigma}\psi \nn\\
&-&\frac{i}{12l^{2}}\theta^{\alpha\beta}\int d^{4}x\;e\;
\varepsilon_{ab}^{\;\;\;\;cd}e_{\alpha}^{a}e_{\beta}^{b}e_{d}^{\sigma}\;\bar{\psi}\gamma_{c}\gamma_{5}
\nabla_{\sigma}\psi\nn \\
&+&\frac{1}{24l} \theta^{\alpha\beta}\int d^{4}x\;e\;\eta_{ab}(\nabla_{\alpha}e_{\mu}^{a})(\nabla_{\beta}e_{\nu}^{b})\;\bar{\psi}\sigma^{\mu\nu}\psi \nn \\  
&-&\frac{i}{24l}\theta^{\alpha\beta}\int d^{4}x\;e\;(\nabla_{\alpha}e_{\mu}^{a})e^{\mu}_{b}\varepsilon_{a}^{\;\;bcd}e_{\beta d}\;\bar{\psi}\gamma_{c}\gamma_{5}\psi \nn\\ 
&-&\frac{1}{24l}\theta^{\alpha\beta}\int d^{4}x\;e\;(\nabla_{\alpha}e_{\mu}^{a})(\nabla_{\beta}e_{\nu}^{b})
(e^{\mu}_{a}e^{\nu}_{c}-e^{\mu}_{c}e^{\nu}_{a})\;\bar{\psi}\sigma^{c}_{\;\;b}\psi\nn\\
&+&\frac{1}{12l^{3}}\theta^{\alpha\beta}\int d^{4}x\;e\;\bar{\psi}\sigma_{\alpha\beta}\psi
\eea

\bea
  \widehat{S}_{4}^{(1)}=&-&\frac{i}{24}\theta^{\alpha\beta}\int d^{4}x\;e\;
  \eta_{ab}(\nabla_{\alpha}e_{\mu}^{a})(\nabla_{\beta}e_{\nu}^{b})\varepsilon^{cdrs}e_{c}^{\mu}e_{d}^{\nu}e_{d}^{\sigma}\;\bar{\psi}\gamma_{r}
 \gamma_{5}\nabla_{\sigma}\psi\nn\\
&-&\frac{i}{24l^{2}}\theta^{\alpha\beta}\int d^{4}x\;e\;
\varepsilon_{ab}^{\;\;\;\;cd}e_{\alpha}^{a}e_{\beta}^{b}e_{d}^{\sigma}\;\bar{\psi}\gamma_{c}\gamma_{5}
\nabla_{\sigma}\psi\nn\\
&+&\frac{1}{24l}\theta^{\alpha\beta}\int d^{4}x\;e\;\eta_{ab}(\nabla_{\alpha}e_{\mu}^{a})(\nabla_{\beta}e_{\nu}^{b})\;\bar{\psi}\sigma^{\mu\nu}\psi\nn\\
             &+&\frac{1}{24l^{3}}\theta^{\alpha\beta}\int d^{4}x\;e\;\bar{\psi}\sigma_{\alpha\beta}\psi
\eea

\bea
\widehat{S}_{5}^{(1)}+\widehat{S}_{6}^{(1)}+\widehat{S}_{7}^{(1)}=&-&\frac{i}{24}\theta^{\alpha\beta}\int d^{4}x\;e\;R_{\alpha\mu}^{\;\;\;\;ab}\varepsilon_{abc}^{\;\;\;\;\;\;d}e_{\beta}^{c}(e^{\mu}_{d}e^{\sigma}_{s}-e^{\mu}_{s}e^{\sigma}_{d})\;\bar{\psi}\gamma^{s}\gamma^{5}\nabla_{\sigma}\psi \nn\\
&-&\frac{i}{4l}\theta^{\alpha\beta}\int d^{4}x\;e\;
T_{\alpha\beta}^{\;\;\;\;a}e^{\sigma}_{a}\;\bar{\psi}\nabla_{\sigma}\psi
+\frac{i}{4l}\theta^{\alpha\beta}\int d^{4}x\;e\;T_{\alpha\mu}^{\;\;\;\;a}e^{\mu}_{a}
\;\bar{\psi}\nabla_{\beta}\psi\nn \\
&-&\frac{1}{12l}\theta^{\alpha\beta}\int d^{4}x\;e\;
T_{\alpha\mu}^{\;\;\;\;a}\varepsilon_{ab}^{\;\;\;\;cd}e_{\beta}^{b}e_{c}^{\mu}e^{\sigma}_{d}\;\bar{\psi}\gamma^{5}\nabla_{\sigma}\psi\nn \\
&+&\frac{i}{12l^{2}}\theta^{\alpha\beta}\int d^{4}x\;e\;
\varepsilon_{abc}^{\;\;\;\;\;\;d}e_{\alpha}^{a}e_{\beta}^{b}e^{\sigma}_{d}\;\bar{\psi}\gamma^{c}\gamma^{5}
\nabla_{\sigma}\psi\nn\\
&-&\frac{1}{16l}\theta^{\alpha\beta}\int d^{4}x\;e\;
R_{\alpha\mu}^{\;\;\;\;ab}\varepsilon_{abc}^{\;\;\;\;\;\;d}e_{\beta}^{c}e^{\mu}_{d}
\;\bar{\psi}\gamma^{5}\psi \nn \\  
&-&\frac{1}{48l}\theta^{\alpha\beta}\int d^{4}x\;e\;
R_{\alpha\beta}^{\;\;\;\;ab}\;\bar{\psi}\sigma_{ab}\psi \nn\\
&-&\frac{1}{24l}\theta^{\alpha\beta}\int d^{4}x\;e\;R_{\alpha\mu}^{\;\;\;\;ab}
e^{\mu}_{a}e^{c}_{\beta}\;\bar{\psi}\sigma_{bc}\psi \nn\\
&+&\frac{i}{24l^{2}}\theta^{\alpha\beta}
\int d^{4}x\;e\;
T_{\alpha\mu}^{\;\;\;\;a}e_{\beta}^{b}
\varepsilon_{ab}^{\;\;\;\;cd}e^{\mu}_{c}\;\bar{\psi}\gamma_{d}\gamma_{5}\psi\nn\\
&+&\frac{1}{8l^{2}}\theta^{\alpha\beta}\int d^{4}x\;e\;
T_{\alpha\beta}^{\;\;\;\;a}\;\bar{\psi}\gamma_{a}\psi \nn\\
&-&\frac{1}{8l^{2}}\theta^{\alpha\beta}\int d^{4}x\;e\;T_{\alpha\mu}^{\;\;\;\;a}e^{\mu}_{a}\;\bar{\psi}\gamma_{\beta}\psi \nn\\
&-&\frac{1}{12l^{3}}\theta^{\alpha\beta}\int d^{4}x\;e\;\bar{\psi}\sigma_{\alpha\beta}\psi
\eea

\bea
  \widehat{S}_{8}^{(1)}=&-&\frac{1}{8}\theta^{\alpha\beta}\int d^{4}x\;e\;R_{\alpha\mu}^{\;\;\;\;ab}e^{\mu}_{a}\;\bar{\psi}\gamma_{b}\nabla_{\beta}\psi 
  -\frac{i}{16}\theta^{\alpha\beta}\int d^{4}x\;e\;
  R_{\alpha \mu}^{\;\;\;\;bc}e^{\mu}_{a}\varepsilon^{a}_{\;\;bcm}\;\bar{\psi}\gamma^{m}\gamma^{5}\nabla_{\beta}\psi\nn\\
  &-&\frac{i}{8l}\theta^{\alpha\beta}\int d^{4}x\;e\;
  T_{\alpha\mu}^{\;\;\;\;a}e^{\mu}_{a}\;\bar{\psi}\nabla_{\beta}\psi
  +\frac{1}{8l}
  \theta^{\alpha\beta}\int d^{4}x\;e\;
  T_{\alpha\mu}^{\;\;\;\;a}e^{\mu}_{b}\;\bar{\psi}\sigma_{a}^{\;\;b}
  \nabla_{\beta}\psi\nn\\
  &-&\frac{3}{8l^{2}}\theta^{\alpha\beta}\int d^{4}x\;e\;\bar{\psi}\gamma_{\alpha}\nabla_{\beta}\psi\nn\\ 
    &-&\frac{1}{16l}\theta^{\alpha\beta}\int d^{4}x\;e\;
  R_{\alpha\mu}^{\;\;\;\;ab}e^{\mu}_{a}e^{c}_{\beta}
  \;\bar{\psi}\sigma_{bc}\psi \nn\\
  &-&\frac{1}{32l}\theta^{\alpha\beta}\int d^{4}x\;e
  \;R_{\alpha\beta}^{\;\;\;\;ab}\;\bar{\psi}\sigma_{ab}\psi 
  -\frac{1}{16l}\theta^{\alpha\beta}\int d^{4}x\;e\;
  R_{\alpha\mu}^{\;\;\;\;ab}e_{\beta a}e^{\mu}_{c}
  \;\bar{\psi}\sigma_{b}^{\;\;c}\psi \nn \\  
  &-&\frac{i}{16l}\theta^{\alpha\beta}\int d^{4}x\;e\;
  R_{\alpha\mu}^{\;\;\;\;ab}e^{\mu}_{a}e_{\beta b}\;\bar{\psi}\psi
  +\frac{1}{32l}\theta^{\alpha\beta}\int d^{4}x\;e\;
  R_{\alpha\mu}^{\;\;\;\;ab}\varepsilon_{abc}^{\;\;\;\;\;\;d}e^{c}_{\beta}e^{\mu}_{d}\;\bar{\psi}\gamma_{5}\psi \nn\\ 
  &-&\frac{1}{16l^{2}}\theta^{\alpha\beta}\int d^{4}x\;e
  \;T_{\alpha\beta}^{\;\;\;\;a}\;\bar{\psi}\gamma_{a}\psi 
  +\frac{1}{16l^{2}}\theta^{\alpha\beta}\int d^{4}x\;e\;
  T_{\alpha\mu}^{\;\;\;\;a}e^{\mu}_{a}\;\bar{\psi}\gamma_{\beta}\psi \nn\\
  &+&\frac{1}{16l^{2}}\theta^{\alpha\beta}
  \int d^{4}x\;e\;T_{\alpha\mu}^{\;\;\;\;a}
  e_{\beta a}\;\bar{\psi}\gamma^{\mu}\psi\nn \\
  &-&\frac{3}{16l^{2}}\theta^{\alpha\beta}\int d^{4}x\;e\;\bar{\psi}\sigma_{\alpha\beta}\psi
\eea

\newpage

\section{Mass terms}
The three mass terms (\ref{mass1}), (\ref{mass2}) and (\ref{mass3}) are 
considered separately after the symmetry breaking. Each of them contains ten 
different terms. In our notation, $\widehat{S}^{(1)}_{i.j}$ is the j-th $(j=1,...10)$ term 
in $\widehat{S}^{(1)}_{m,i}$ $(i=1,2,3)$. The terms are listed according to the order of 
appearance in (\ref{mass1}), (\ref{mass2}) and (\ref{mass3}): 
\\[0.2cm]
\bea
  \widehat{S}^{(1)}_{1.1}=&-&6ic_{1}\left(m-\frac{2}{l}\right)\theta^{\alpha\beta}\int d^{4}x\;e\;(\nabla_{\alpha}e_{\mu}^{a})e^{\mu}_{a}\;\bar{\psi}\nabla_{\beta}\psi \nn \\
&+&3c_{1}\left(\frac{m}{l}-\frac{2}{l^{2}}\right)\theta^{\alpha\beta}\int d^{4}x\;e\;(\nabla_{\alpha}e_{\mu}^{a})e^{\mu}_{a}\;\bar{\psi}\gamma_{\beta}\psi \\[0.2cm]
  \widehat{S}^{(1)}_{1.2}=&+&\frac{3c_{1}}{4}\left(m-\frac{2}{l}\right)\theta^{\alpha\beta}\int d^{4}x\;e\;R_{\alpha\beta}^{\;\;\;\;ab}\;\bar{\psi}\sigma_{ab}\psi \nn \\
     &-&\frac{3c_{1}}{2}\left(\frac{m}{l}-\frac{2}{l^{2}}\right)\theta^{\alpha\beta}\int d^{4}x\;e\;T_{\alpha\beta}^{\;\;\;\;a}\;\bar{\psi}\gamma_{a}\psi \nn \\
     &-&\frac{3c_{1}}{2}\left(\frac{m}{l^{2}}-\frac{2}{l^{3}}\right)\theta^{\alpha\beta}
    \int d^{4}x\;e\;\bar{\psi}\sigma_{\alpha\beta}\psi \\[0.2cm]
  \widehat{S}^{(1)}_{1.3}=&-&6c_{1}\left(\frac{m}{l}-\frac{2}{l^{2}}\right)\theta^{\alpha\beta}\int d^{4}x\;e\;(\nabla_{\alpha}e_{\mu}^{a})e^{\mu}_{a}\;\bar{\psi}\gamma_{\beta}\psi \nn \\
 &-&6c_{1}\left(\frac{m}{l^{2}}-\frac{2}{l^{3}}\right)\theta^{\alpha\beta}\int d^{4}x\;e\;\bar{\psi}\sigma_{\alpha\beta}\psi \\ [0.2cm]
  \widehat{S}^{(1)}_{1.4}=&-&2c_{1}\left(m-\frac{2}{l}\right)\theta^{\alpha\beta}\int d^{4}x\;e\;(\nabla_{\alpha}e_{\mu}^{a})(\nabla_{\beta}e_{\nu}^{b})
  (e^{\mu}_{a}e^{\nu}_{c}-e^{\mu}_{c}e^{\nu}_{a})\;\bar{\psi}\sigma^{c}_{\;\;b}\psi \nn \\
                &+&2c_{1}\left(\frac{m}{l^{2}}-\frac{2}{l^{3}}\right)\theta^{\alpha\beta}
  \int d^{4}x\;e\;\bar{\psi}\sigma_{\alpha\beta}\psi\\[0.2cm]
\widehat{S}^{(1)}_{1.5}=&+&\frac{c_{1}}{2}\left(m-\frac{2}{l}\right)\theta^{\alpha\beta}
\int d^{4}x\;e\;\eta_{ab}(\nabla_{\alpha}e_{\mu}^{a})
(\nabla_{\beta}e_{\nu}^{b})\;\bar{\psi}\sigma^{\mu\nu}\psi \nn \\
    &+&\frac{c_{1}}{2}\left(\frac{m}{l^{2}}-\frac{2}{l^{3}}\right)\theta^{\alpha\beta}
    \int d^{4}x\;e\;\bar{\psi}\sigma_{\alpha\beta}\psi\\[0.2cm]
  \widehat{S}^{(1)}_{1.6}=&+&\frac{c_{1}}{2}\left(m-\frac{2}{l}\right)\theta^{\alpha\beta}
  \int d^{4}x\;e\;\eta_{ab}(\nabla_{\alpha}e_{\mu}^{a})
  (\nabla_{\beta}e_{\nu}^{b})\;\bar{\psi}\sigma^{\mu\nu}\psi \nn \\
   &+&\frac{c_{1}}{2}\left(\frac{m}{l^{2}}-\frac{2}{l^{3}}\right)\theta^{\alpha\beta}\int d^{4}x\;e\;\bar{\psi}\sigma_{\alpha\beta}\psi
\eea

\newpage

\bea
  \widehat{S}^{(1)}_{1.7}=&-&\frac{3c_{1}}{4}\mass\theta^{\alpha\beta}\int d^{4}x\;e\;R_{\alpha\beta}^{\;\;\;\;ab}\;\bar{\psi}\sigma_{ab}\psi \nn\\
&-&\frac{3c_{1}}{2}\mass\theta^{\alpha\beta}\int d^{4}x\;e\;R_{\alpha\mu}^{\;\;\;\;ab}e^{\mu}_{a}e_{\beta}^{c}\;\bar{\psi}\sigma_{bc}\psi \nn \\ 
&+&\frac{3c_{1}}{2}\massl\theta^{\alpha\beta}\int d^{4}x\;e
\;T_{\alpha\beta}^{\;\;\;\;a}\;\bar{\psi}\gamma_{a}\psi \nn\\
&-&\frac{3c_{1}}{2}\massl\theta^{\alpha\beta}\int d^{4}x\;e
\;T_{\alpha\mu}^{\;\;\;\;a}e^{\mu}_{a}\;\bar{\psi}\gamma_{\beta}\psi \nn \\
&+&\frac{9c_{1}}{2}\massll\theta^{\alpha\beta}\int d^{4}x\;e\;\bar{\psi}\sigma_{\alpha\beta}\psi \\[0.2cm]
\widehat{S}^{(1)}_{1.8}=&+&\frac{c_{1}}{4}\mass\theta^{\alpha\beta}\int d^{4}x\;e\;R_{\alpha\beta}^{\;\;\;\;ab}\;\bar{\psi}\sigma_{ab}\psi \nn\\
&+&\frac{c_{1}}{2}\mass\theta^{\alpha\beta}\int d^{4}x\;e\;R_{\alpha\mu}^{\;\;\;\;ab}e^{\mu}_{a}e_{\beta}^{c}\;\bar{\psi}\sigma_{bc}\psi \nn \\
&+&\frac{c_{1}}{2}\massl\theta^{\alpha\beta}\int d^{4}x\;e
\;T_{\alpha\beta}^{\;\;\;\;a}\;\bar{\psi}\gamma_{a}\psi \nn\\ 
&-&\frac{c_{1}}{2}\massl\theta^{\alpha\beta}\int d^{4}x\;e
\;T_{\alpha\mu}^{\;\;\;\;a}e^{\mu}_{a}\;\bar{\psi}\gamma_{\beta}\psi \nn \\
&+&c_{1}\massll\theta^{\alpha\beta}\int d^{4}x\;e\;\bar{\psi}\sigma_{\alpha\beta}\psi \\[0.2cm]
  \widehat{S}^{(1)}_{1.9}=&+&\frac{c_{1}}{4}\mass\theta^{\alpha\beta}\int d^{4}x\;e\;R_{\alpha\beta}^{\;\;\;\;ab}\;\bar{\psi}\sigma_{ab}\psi \nn\\
&+&\frac{c_{1}}{2}\mass\theta^{\alpha\beta}\int d^{4}x\;e\;R_{\alpha\mu}^{\;\;\;\;ab}e^{\mu}_{a}e_{\beta}^{c}\;\bar{\psi}\sigma_{bc}\psi \nn \\
&-&\frac{c_{1}}{2}\massl\theta^{\alpha\beta}\int d^{4}x\;e
\;T_{\alpha\beta}^{\;\;\;\;a}\;\bar{\psi}\gamma_{a}\psi \nn\\
&+&\frac{c_{1}}{2}\massl\theta^{\alpha\beta}\int d^{4}x\;e
\;T_{\alpha\mu}^{\;\;\;\;a}e^{\mu}_{a}\;\bar{\psi}\gamma_{\beta}\psi \nn \\
&+&c_{1}\massll\theta^{\alpha\beta}\int d^{4}x\;e\;\bar{\psi}\sigma_{\alpha\beta}\psi \\[0.2cm]
\widehat{S}^{(1)}_{1.10}=&-&\frac{3c_{1}}{4}\mass\theta^{\alpha\beta}\int d^{4}x\;e\;R_{\alpha\beta}^{\;\;\;\;ab}\;\bar{\psi}\sigma_{ab}\psi \nn\\ 
&-&\frac{3c_{1}}{2}\mass\theta^{\alpha\beta}\int d^{4}x\;e
\;R_{\alpha\mu}^{\;\;\;\;ab}e^{\mu}_{a}e_{\beta}^{c}\;\bar{\psi}\sigma_{bc}\psi \nn \\
&-&\frac{3c_{1}}{2}\massl\theta^{\alpha\beta}
\int d^{4}x\;e\;T_{\alpha\beta}^{\;\;\;\;a}\;\bar{\psi}\gamma_{a}\psi \nn\\
&+&\frac{3c_{1}}{2}\massl\theta^{\alpha\beta}
\int d^{4}x\;e\;T_{\alpha\mu}^{\;\;\;\;a}e^{\mu}_{a}\;\bar{\psi}\gamma_{\beta}\psi \nn \\
&-&3c_{1}\massll\theta^{\alpha\beta}\int d^{4}x\;e\;\bar{\psi}\sigma_{\alpha\beta}\psi 
\eea

\bea  
 \widehat{S}^{(1)}_{2.1}=&+&6ic_{2}\mass\theta^{\alpha\beta}\int d^{4}x\;e\;(\nabla_{\alpha}e_{\mu}^{a})e^{\mu}_{a}\;\bar{\psi}\nabla_{\beta}\psi \nn \\
&-&3c_{2}\massl\theta^{\alpha\beta}\int d^{4}x\;e\;(\nabla_{\alpha}e_{\mu}^{a})e^{\mu}_{a}\;\bar{\psi}\gamma_{\beta}\psi \\[0.2cm]
  \widehat{S}^{(1)}_{2.2}=&-&\frac{3c_{2}}{4}\mass\theta^{\alpha\beta}\int d^{4}x\;e\;R_{\alpha\beta}^{\;\;\;\;ab}\;\bar{\psi}\sigma_{ab}\psi \nn \\ 
                &+&\frac{3c_{2}}{2}\massl\theta^{\alpha\beta}\int d^{4}x\;e\;T_{\alpha\beta}^{\;\;\;\;a}\;\bar{\psi}\gamma_{a}\psi \nn \\
                &+&\frac{3c_{2}}{2}\massll\theta^{\alpha\beta}
     \int d^{4}x\;e\;\bar{\psi}\sigma_{\alpha\beta}\psi \\[0.2cm]
\widehat{S}^{(1)}_{2.3}=&+&\frac{3c_{2}}{2}\mass\theta^{\alpha\beta}\int d^{4}x\;e\;(\nabla_{\alpha}e_{\mu}^{a})(\nabla_{\beta}e_{\nu}^{b})(e^{\mu}_{b}e^{\nu}_{c}-e^{\mu}_{c}e^{\nu}_{b})\;\bar{\psi}\sigma^{c}_{\;\;a}\psi \nn \\
 &+&\frac{3c_{2}}{2}\massl\theta^{\alpha\beta}\int d^{4}x\;e\;(\nabla_{\alpha}e_{\mu}^{a})e^{\mu}_{a}\;\bar{\psi}\gamma_{\beta}\psi \nn \\
 &-&\frac{3c_{2}}{2}\massl\theta^{\alpha\beta}\int d^{4}x\;e\;(\nabla_{\alpha}e^{a}_{\mu})e^{\nu}_{a}g_{\nu\beta}\;\bar{\psi}\gamma^{\mu}\psi \nn \\
    &+&\frac{3c_{2}}{4}\massl\theta^{\alpha\beta}\int d^{4}x\;e\;T_{\alpha\beta}^{\;\;\;\;a}\;\bar{\psi}\gamma_{a}\psi \\[0.2cm]
\widehat{S}^{(1)}_{2.4}=&+&\frac{3c_{2}}{2}\massl\theta^{\alpha\beta}
\int d^{4}x\;e\;\eta_{ab}(\nabla_{\alpha}e_{\mu}^{a})e_{\beta}^{b}\;\bar{\psi}\gamma^{\mu}\psi \nn \\
&+&\frac{3c_{2}}{2}\massl\theta^{\alpha\beta}
\int d^{4}x\;e\;\eta_{ab}(\nabla_{\alpha}e_{\mu}^{a})e^{\mu}_{a}\;\bar{\psi}\gamma_{\beta}\psi \nn \\
&+&\frac{3c_{2}}{4}\massl\theta^{\alpha\beta}\int d^{4}x\;e
\;T_{\alpha\beta}^{\;\;\;\;a}\;\bar{\psi}\gamma_{a}\psi \nn \\
&-&\frac{3c_{2}}{2}\massll\theta^{\alpha\beta}\int d^{4}x\;e\;\bar{\psi}\sigma_{\alpha\beta}\psi \\[0.2cm]
\widehat{S}^{(1)}_{2.5}=&-&\frac{c_{2}}{2}\mass\theta^{\alpha\beta}\int d^{4}x\;e
\;\eta_{ab}(\nabla_{\alpha}e_{\mu}^{a})(\nabla_{\beta}e_{\nu}^{b})\;\bar{\psi}\sigma^{\mu\nu}\psi \nn \\
                &+&\frac{c_{2}}{2}\mass\theta^{\alpha\beta}
                \int d^{4}x\;e\;(\nabla_{\alpha}e_{\mu}^{a})(\nabla_{\beta}e_{\nu}^{b})(e^{\mu}_{b}e^{\nu}_{c}-e^{\mu}_{c}e^{\nu}_{b})\;\bar{\psi}\sigma^{c}_{\;\;a}\psi \nn \\
                &-&c_{2}\massll\theta^{\alpha\beta}
    \int d^{4}x\;e\;\bar{\psi}\sigma_{\alpha\beta}\psi \\[0.2cm]
\widehat{S}^{(1)}_{2.6}=&-&\frac{c_{2}}{2}\mass\theta^{\alpha\beta}
\int d^{4}x\;e\;\eta_{ab}(\nabla_{\alpha}e_{\mu}^{a})
(\nabla_{\beta}e_{\nu}^{b})\;\bar{\psi}\sigma^{\mu\nu}\psi \nn \\
                &-&\frac{c_{2}}{2}\massll\theta^{\alpha\beta}
    \int d^{4}x\;e\;\bar{\psi}\sigma_{\alpha\beta}\psi
\eea

\bea
  \widehat{S}^{(1)}_{2.7}=&+&\frac{3c_{2}}{4}\mass\theta^{\alpha\beta}\int d^{4}x\;e\;R_{\alpha\beta}^{\;\;\;\;ab}\;\bar{\psi}\sigma_{ab}\psi \nn\\
&+&\frac{3c_{2}}{2}\mass\theta^{\alpha\beta}\int d^{4}x\;e
\;R_{\alpha\mu}^{\;\;\;\;ab}e^{\mu}_{a}e_{\beta}^{c}\;\bar{\psi}\sigma_{bc}\psi \nn \\ 
&-&\frac{3c_{2}}{2}\massl\theta^{\alpha\beta}
\int d^{4}x\;e\;T_{\alpha\beta}^{\;\;\;\;a}\;\bar{\psi}\gamma_{a}\psi \nn\\ 
&+&\frac{3c_{2}}{2}\massl\theta^{\alpha\beta}
\int d^{4}x\;e\;T_{\alpha\mu}^{\;\;\;\;a}e^{\mu}_{a}\;\bar{\psi}\gamma_{\beta}\psi \nn \\
&-&\frac{9c_{2}}{2}\massll\theta^{\alpha\beta}
\int d^{4}x\;e\;\bar{\psi}\sigma_{\alpha\beta}\psi \\[0.2cm] 
  \widehat{S}^{(1)}_{2.8}=&-&\frac{c_{2}}{4}\mass\theta^{\alpha\beta}\int d^{4}x\;e\;R_{\alpha\beta}^{\;\;\;\;ab}\;\bar{\psi}\sigma_{ab}\psi \nn\\
&-&\frac{c_{2}}{2}\mass\theta^{\alpha\beta}\int d^{4}x\;e\;R_{\alpha\mu}^{\;\;\;\;ab}e^{\mu}_{a}e_{\beta}^{c}\;\bar{\psi}\sigma_{bc}\psi \nn \\
&-&\frac{c_{2}}{2}\massl\theta^{\alpha\beta}
\int d^{4}x\;e\;T_{\alpha\beta}^{\;\;\;\;a}\;\bar{\psi}\gamma_{a}\psi \nn\\
&+&\frac{c_{2}}{2}\massl\theta^{\alpha\beta}
\int d^{4}x\;e\;T_{\alpha\mu}^{\;\;\;\;a}e^{\mu}_{a}\;\bar{\psi}\gamma_{\beta}\psi \nn \\
&-&c_{2}\massll\theta^{\alpha\beta}\int d^{4}x\;e\;\bar{\psi}\sigma_{\alpha\beta}\psi\\[0.2cm]
  \widehat{S}^{(1)}_{2.9}=&-&\frac{c_{2}}{4}\mass\theta^{\alpha\beta}\int d^{4}x\;e\;R_{\alpha\beta}^{\;\;\;\;ab}\;\bar{\psi}\sigma_{ab}\psi \nn\\
&-&\frac{c_{2}}{2}\mass\theta^{\alpha\beta}\int d^{4}x\;e
\;R_{\alpha\mu}^{\;\;\;\;ab}e^{\mu}_{a}e_{\beta}^{c}\;\bar{\psi}\sigma_{bc}\psi \nn \\  
&+&\frac{c_{2}}{2}\massl\theta^{\alpha\beta}
\int d^{4}x\;e\;T_{\alpha\beta}^{\;\;\;\;a}\;\bar{\psi}\gamma_{a}\psi \nn\\ 
&-&\frac{c_{2}}{2}\massl\theta^{\alpha\beta}
\int d^{4}x\;e\;T_{\alpha\mu}^{\;\;\;\;a}e^{\mu}_{a}\;\bar{\psi}\gamma_{\beta}\psi \nn \\ 
&-&c_{2}\massll\theta^{\alpha\beta}\int d^{4}x\;e\;\bar{\psi}\sigma_{\alpha\beta}\psi   \\[0.2cm]
  \widehat{S}^{(1)}_{2.10}=&+&\frac{3c_{2}}{4}\mass\theta^{\alpha\beta}\int d^{4}x\;e\;R_{\alpha\beta}^{\;\;\;\;ab}\;\bar{\psi}\sigma_{ab}\psi \nn\\
&+&\frac{3c_{2}}{2}\mass\theta^{\alpha\beta}\int d^{4}x\;e
\;R_{\alpha\mu}^{\;\;\;\;ab}e^{\mu}_{a}e_{\beta}^{c}\;\bar{\psi}\sigma_{bc}\psi \nn \\
&-&\frac{3c_{2}}{2}\massl\theta^{\alpha\beta}
\int d^{4}x\;e\;T_{\alpha\beta}^{\;\;\;\;a}\;\bar{\psi}\gamma_{a}\psi \nn\\
&+&\frac{3c_{2}}{2}\massl\theta^{\alpha\beta}
\int d^{4}x\;e\;T_{\alpha\mu}^{\;\;\;\;a}e^{\mu}_{a}\;\bar{\psi}\gamma_{\beta}\psi \nn \\
&+&3c_{2}\massll\theta^{\alpha\beta}\int d^{4}x\;e\;\bar{\psi}\sigma_{\alpha\beta}\psi
\eea

\bea
  \widehat{S}^{(1)}_{3.1}=&-&6ic_{3}\mass\theta^{\alpha\beta}\int d^{4}x\;e\;(\nabla_{\alpha}e_{\mu}^{a})e^{\mu}_{a}\;\bar{\psi}
  \nabla_{\beta}\psi \nn \\
&+&3c_{3}\massl\theta^{\alpha\beta}\int d^{4}x\;e\;(\nabla_{\alpha}e_{\mu}^{a})e^{\mu}_{a}\;\bar{\psi}\gamma_{\beta}\psi \\[0.2cm]
  \widehat{S}^{(1)}_{3.2}=&+&\frac{3c_{3}}{4}\mass\theta^{\alpha\beta}\int d^{4}x\;e\;R_{\alpha\beta}^{\;\;\;\;ab}\;\bar{\psi}\sigma_{ab}\psi \nn \\
                &-&\frac{3c_{3}}{2}\massl\theta^{\alpha\beta}\int d^{4}x\;e\;T_{\alpha\beta}^{\;\;\;\;a}\;\bar{\psi}\gamma_{a}\psi \nn \\
                &-&\frac{3c_{3}}{2}\massll\theta^{\alpha\beta}
                \int d^{4}x\;e\;\bar{\psi}\sigma_{\alpha\beta}\psi \\[0.2cm]
  \widehat{S}^{(1)}_{3.3}=&-&2c_{3}\mass\theta^{\alpha\beta}\int d^{4}x\;e\;(\nabla_{\alpha}e_{\mu}^{a})(\nabla_{\beta}e_{\nu}^{b})
  (e^{\mu}_{b}e^{\nu}_{c}-e^{\mu}_{c}e^{\nu}_{b})\;\bar{\psi}\sigma^{c}_{\;\;a}\psi \nn \\
               &-&3c_{3}\massl\theta^{\alpha\beta}\int d^{4}x\;e\;(\nabla_{\alpha}e_{\mu}^{a})e^{\mu}_{a}\;\bar{\psi}\gamma_{\beta}\psi \nn \\
               &+&3c_{3}\massl\theta^{\alpha\beta}\int d^{4}x\;e\;(\nabla_{\alpha}e^{a}_{\mu})e^{\nu}_{a}g_{\nu\beta}\;\bar{\psi}\gamma^{\mu}\psi \nn \\
&-&c_{3}\massl\theta^{\alpha\beta}\int d^{4}x\;e
\;T_{\alpha\beta}^{\;\;\;\;a}\;\bar{\psi}\gamma_{a}\psi \nn \\
&-&c_{3}\massll\theta^{\alpha\beta}\int d^{4}x\;e\;\bar{\psi}\sigma_{\alpha\beta}\psi \\[0.2cm]
\widehat{S}^{(1)}_{3.4}=&-&2c_{3}\massl\theta^{\alpha\beta}
\int d^{4}x\;e\;\eta_{ab}(\nabla_{\alpha}e_{\mu}^{a})e_{\beta}^{b}\;\bar{\psi}\gamma^{\mu}\psi \nn \\
  &+&c_{3}\massl\theta^{\alpha\beta}\int d^{4}x\;e\;(\nabla_{\alpha}e_{\mu}^{a})e^{\mu}_{a}\;\bar{\psi}\gamma_{\beta}\psi \nn \\
  &-&c_{3}\massl\theta^{\alpha\beta}\int d^{4}x\;e
  \;T_{\alpha\beta}^{\;\;\;\;a}\;\bar{\psi}\gamma_{a}\psi \nn \\
                &+&c_{3}\massll\theta^{\alpha\beta}
                \int d^{4}x\;e\;\bar{\psi}\sigma_{\alpha\beta}\psi \\[0.2cm]
\widehat{S}^{(1)}_{3.5}=&+&\frac{c_{3}}{2}\mass\theta^{\alpha\beta}
\int d^{4}x\;e\;\eta_{ab}(\nabla_{\alpha}e_{\mu}^{a})
(\nabla_{\beta}e_{\nu}^{b})\;\bar{\psi}\sigma^{\mu\nu}\psi \nn \\
                &+&\frac{c_{3}}{2}\massll\theta^{\alpha\beta}
                \int d^{4}x\;e\;\bar{\psi}\sigma_{\alpha\beta}\psi\\[0.2cm]
\widehat{S}^{(1)}_{3.6}=&+&\frac{c_{3}}{2}\mass\theta^{\alpha\beta}
\int d^{4}x\;e\;\eta_{ab}(\nabla_{\alpha}e_{\mu}^{a})(\nabla_{\beta}e_{\nu}^{b})\;\bar{\psi}\sigma^{\mu\nu}\psi \nn \\
                &+&\frac{c_{3}}{2}\massll\theta^{\alpha\beta}
                \int d^{4}x\;e\;\bar{\psi}\sigma_{\alpha\beta}\psi
\eea

\bea
  \widehat{S}^{(1)}_{3.7}=&-&\frac{3c_{3}}{4}\mass\theta^{\alpha\beta}\int d^{4}x\;e\;R_{\alpha\beta}^{\;\;\;\;ab}\;\bar{\psi}\sigma_{ab}\psi \nn\\
&-&\frac{3c_{3}}{2}\mass\theta^{\alpha\beta}\int d^{4}x\;e
\;R_{\alpha\mu}^{\;\;\;\;ab}e^{\mu}_{a}e_{\beta}^{c}\;\bar{\psi}\sigma_{bc}\psi \nn \\  
&+&\frac{3c_{3}}{2}\massl\theta^{\alpha\beta}\int d^{4}x\;e
\;T_{\alpha\beta}^{\;\;\;\;a}\;\bar{\psi}\gamma_{a}\psi \nn\\ 
&-&\frac{3c_{3}}{2}\massl\theta^{\alpha\beta}\int d^{4}x\;e
\;T_{\alpha\mu}^{\;\;\;\;a}e^{\mu}_{a}\;\bar{\psi}\gamma_{\beta}\psi \nn \\
&+&\frac{9c_{3}}{2}\massll\theta^{\alpha\beta}\int d^{4}x\;e\;\bar{\psi}\sigma_{\alpha\beta}\psi \\[0.2cm]  
  \widehat{S}^{(1)}_{3.8}=&+&\frac{c_{3}}{4}\mass\theta^{\alpha\beta}\int d^{4}x\;e\;R_{\alpha\beta}^{\;\;\;\;ab}\;\bar{\psi}\sigma_{ab}\psi \nn\\ 
&+&\frac{c_{3}}{2}\mass\theta^{\alpha\beta}\int d^{4}x\;e
\;R_{\alpha\mu}^{\;\;\;\;ab}e^{\mu}_{a}e_{\beta}^{c}\;\bar{\psi}
\sigma_{bc}\psi \nn \\
&+&\frac{c_{3}}{2}\massl\theta^{\alpha\beta}\int d^{4}x\;e
\;T_{\alpha\beta}^{\;\;\;\;a}\;\bar{\psi}\gamma_{a}\psi \nn\\
&-&\frac{c_{3}}{2}\massl\theta^{\alpha\beta}\int d^{4}x\;e
\;T_{\alpha\mu}^{\;\;\;\;a}e^{\mu}_{a}\;\bar{\psi}\gamma_{\beta}\psi \nn \\
&+&c_{3}\massll\theta^{\alpha\beta}\int d^{4}x\;e\;\bar{\psi}\sigma_{\alpha\beta}\psi\\[0.2cm]
  \widehat{S}^{(1)}_{3.9}=&+&\frac{c_{3}}{4}\mass\theta^{\alpha\beta}\int d^{4}x\;e\;R_{\alpha\beta}^{\;\;\;\;ab}\;\bar{\psi}\sigma_{ab}\psi \nn\\
&+&\frac{c_{3}}{2}\mass\theta^{\alpha\beta}\int d^{4}x\;e
\;R_{\alpha\mu}^{\;\;\;\;ab}e^{\mu}_{a}e_{\beta}^{c}\;\bar{\psi}\sigma_{bc}\psi \nn \\  
&+&\frac{c_{3}}{2}\massl\theta^{\alpha\beta}\int d^{4}x\;e
\;T_{\alpha\beta}^{\;\;\;\;a}\;\bar{\psi}\gamma_{a}\psi \nn\\ 
&-&\frac{c_{3}}{2}\massl\theta^{\alpha\beta}\int d^{4}x\;e
\;T_{\alpha\mu}^{\;\;\;\;a}e^{\mu}_{a}\;\bar{\psi}\gamma_{\beta}\psi \nn \\  
&+&c_{3}\massll\theta^{\alpha\beta}\int d^{4}x\;e\;\bar{\psi}\sigma_{\alpha\beta}\psi  \\[0.2cm]
  \widehat{S}^{(1)}_{3.10}=&-&\frac{3c_{3}}{4}\mass\theta^{\alpha\beta}\int d^{4}x\;e\;R_{\alpha\beta}^{\;\;\;\;ab}\;\bar{\psi}\sigma_{ab}\psi \nn\\ 
&-&\frac{3c_{3}}{2}\mass\theta^{\alpha\beta}\int d^{4}x\;e\;R_{\alpha\mu}^{\;\;\;\;ab}e^{\mu}_{a}e_{\beta}^{c}\;\bar{\psi}\sigma_{bc}\psi \nn \\
&+&\frac{3c_{3}}{2}\massl\theta^{\alpha\beta}\int d^{4}x\;e\;T_{\alpha\beta}^{\;\;\;\;a}\;\bar{\psi}\gamma_{a}\psi \nn\\
&-&\frac{3c_{3}}{2}\massl\theta^{\alpha\beta}\int d^{4}x\;e\;T_{\alpha\mu}^{\;\;\;\;a}e^{\mu}_{a}\;\bar{\psi}\gamma_{\beta}\psi \nn \\
&-&3c_{3}\massll\theta^{\alpha\beta}\int d^{4}x\;e\;\bar{\psi}\sigma_{\alpha\beta}\psi
\eea

\begin{align}
   \widehat{S}^{(1)}_{m,1}=\sum_{j=1}^{10}\widehat{S}^{(1)}_{1.j}=&-6ic_{1}\mass\theta^{\alpha\beta}\int 
d^{4}x\;e\;(\nabla_{\alpha}e_{\mu}^{a})e^{\mu}_{a}\;\bar{\psi}\nabla_{\beta}\psi 
\nn \\
&+c_{1}\mass\theta^{\alpha\beta}\int d^{4}x\;e\;\eta_{ab}(\nabla_{\alpha}e_{\mu}^{a})(\nabla_{\beta}e_{\nu}^{b})\;\bar{\psi}\sigma^{\mu\nu}\psi \nn\\
&-2c_{1}\mass\theta^{\alpha\beta}\int d^{4}x\;e\;(\nabla_{\alpha}e_{\mu}^{a})(\nabla_{\beta}e_{\nu}^{b})(e^{\mu}_{a}e^{\nu}_{c}-e^{\mu}_{c}e^{\nu}_{a})\;\bar{\psi}\sigma^{c}_{\;\;b}\psi \nn\\
&-3c_{1}\massl\theta^{\alpha\beta}\int d^{4}x\;e\;(\nabla_{\alpha}e_{\mu}^{a})e^{\mu}_{a}\;\bar{\psi}\gamma_{\beta}\psi \nn\\
&-\frac{c_{1}}{4}\mass\theta^{\alpha\beta}\int d^{4}x\;e\;R_{\alpha\beta}^{\;\;\;\;ab}\;\bar{\psi}\sigma_{ab}\psi \nn\\
                              &-2c_{1}\mass\theta^{\alpha\beta}\int d^{4}x\;e\;R_{\alpha\mu}^{\;\;\;\;ab}e^{\mu}_{a}e_{\beta}^{c}\;\bar{\psi}\sigma_{bc}\psi \nn \\
&-\frac{3c_{1}}{2}\massl\theta^{\alpha\beta}\int d^{4}x\;e\;T_{\alpha\beta}^{\;\;\;\;a}\;\bar{\psi}\gamma_{a}\psi \nn\\
&-c_{1}\massll\theta^{\alpha\beta}\int d^{4}x\;e\;\bar{\psi}\sigma_{\alpha\beta}\psi
\end{align}

\begin{align}
   \widehat{S}^{(1)}_{m,2}=\sum_{j=1}^{10}\widehat{S}^{(1)}_{2.j}=&+6ic_{2}\mass\theta^{\alpha\beta}\int 
d^{4}x\;e\;(\nabla_{\alpha}e_{\mu}^{a})e^{\mu}_{a}\;\bar{\psi}\nabla_{\beta}\psi 
\nn\\
&-c_{2}\mass\theta^{\alpha\beta}\int d^{4}x\;e\;\eta_{ab}(\nabla_{\alpha}e_{\mu}^{a})(\nabla_{\beta}e_{\nu}^{b})\;\bar{\psi}\sigma^{\mu\nu}\psi \nn\\
&+2c_{2}\mass\theta^{\alpha\beta}\int d^{4}x\;e\;(\nabla_{\alpha}e_{\mu}^{a})(\nabla_{\beta}e_{\nu}^{b})(e^{\mu}_{a}e^{\nu}_{c}-e^{\mu}_{c}e^{\nu}_{a})\;\bar{\psi}\sigma^{c}_{\;\;b}\psi \nn\\
&+\frac{c_{2}}{4}\mass\theta^{\alpha\beta}\int d^{4}x\;e\;R_{\alpha\beta}^{\;\;\;\;ab}\;\bar{\psi}\sigma_{ab}\psi \nn\\
&+2c_{2}\mass\theta^{\alpha\beta}\int d^{4}x\;e\;R_{\alpha\mu}^{\;\;\;\;ab}e^{\mu}_{a}e_{\beta}^{c}\;\bar{\psi}\sigma_{bc}\psi \nn\\
&+3c_{2}\massl\theta^{\alpha\beta}\int d^{4}x\;e\;T_{\alpha\mu}^{\;\;\;\;a}e^{\mu}_{a}\;\bar{\psi}\gamma_{\beta}\psi \nn\\
&-5c_{2}\massll\theta^{\alpha\beta}\int d^{4}x\;e\;\bar{\psi}\sigma_{\alpha\beta}\psi
\end{align}

\begin{align}
   \widehat{S}^{(1)}_{m,3}=\sum_{j=1}^{10}\widehat{S}^{(1)}_{3.j}=&-6ic_{3}\mass\theta^{\alpha\beta}\int 
d^{4}x\;e\;(\nabla_{\alpha}e_{\mu}^{a})e^{\mu}_{a}\;\bar{\psi}\nabla_{\beta}\psi 
\nn\\
&+c_{3}\mass\theta^{\alpha\beta}\int d^{4}x\;e\;\eta_{ab}(\nabla_{\alpha}e_{\mu}^{a})(\nabla_{\beta}e_{\nu}^{b})\;\bar{\psi}\sigma^{\mu\nu}\psi \nn\\
&-2c_{3}\mass\theta^{\alpha\beta}\int d^{4}x\;e\;(\nabla_{\alpha}e_{\mu}^{a})(\nabla_{\beta}e_{\nu}^{b})(e^{\mu}_{a}e^{\nu}_{c}-e^{\mu}_{c}e^{\nu}_{a})\;\bar{\psi}\sigma^{c}_{\;\;b}\psi \nn\\
&+c_{3}\massl\theta^{\alpha\beta}\int d^{4}x\;e\;(\nabla_{\alpha}e_{\mu}^{a})e^{\mu}_{a}\;\bar{\psi}\gamma_{\beta}\psi \nn\\
&-\frac{c_{3}}{4}\mass\theta^{\alpha\beta}\int d^{4}x\;e\;R_{\alpha\beta}^{\;\;\;\;ab}\;\bar{\psi}\sigma_{ab}\psi \nn\\
&-2c_{3}\mass\theta^{\alpha\beta}\int d^{4}x\;e\;R_{\alpha\mu}^{\;\;\;\;ab}e^{\mu}_{a}e_{\beta}^{c}\;\bar{\psi}\sigma_{bc}\psi \nn\\
&+\frac{1}{2}c_{3}\massl\theta^{\alpha\beta}\int d^{4}x\;e\;T_{\alpha\beta}^{\;\;\;\;a}\;\bar{\psi}\gamma_{a}\psi \nn\\
&-4c_{3}\massl\theta^{\alpha\beta}\int d^{4}x\;e\;T_{\alpha\mu}^{\;\;\;\;a}e^{\mu}_{a}\;\bar{\psi}\gamma_{\beta}\psi \nn\\
&+3c_{3}\massll\theta^{\alpha\beta}\int d^{4}x\;e\;\bar{\psi}\sigma_{\alpha\beta}\psi
\end{align}

\end{document}